\documentclass[nojss]{jss}

\usepackage[utf8]{inputenc}

\author{
Alex Stringer\\University of Toronto
}
\title{Implementing Approximate Bayesian Inference using Adaptive Quadrature: the \pkg{aghq}
Package}

\Plainauthor{Alex Stringer}
\Plaintitle{Implementing Adaptive Quadrature for Bayesian Inference: the aghq
Package}
\Shorttitle{\pkg{aghq}: Adaptive Quadrature for Bayesian Inference}

\Abstract{
The \pkg{aghq} package for implementing approximate Bayesian inference
using adaptive quadrature is introduced. The method and
software are described, and use of the package in making approximate Bayesian inferences in several challenging low- and
high-dimensional models is illustrated. Examples include an infectious disease model; an astrostatistical model for
estimating the mass of the Milky Way; two examples in non-Gaussian model-based geostatistics including
one incorporating zero-inflation which is not easily fit using other methods; and a
model for zero-inflated, overdispersed count data. The \pkg{aghq} package is especially compatible
with the popular \pkg{TMB} interface for automatic differentiation and Laplace approximation,
and existing users of that software can make approximate Bayesian inferences with \pkg{aghq}
using very little additional code. The \pkg{aghq} package is available from \texttt{CRAN} and
complete code for all examples in this paper can be found
at \href{https://github.com/awstringer1/aghq-software-paper-code}{https://github.com/awstringer1/aghq-software-paper-code}.
}

\Plainkeywords{bayesian, quadrature, optimization, spatial}

\Address{
University of Toronto, Toronto, Canada \\
Centre for Global Health Research, St. Michael's Hospital, Toronto, Canada
}

\usepackage{preamble}

\begin{document}

\hypertarget{introduction}{%
\section{Introduction}\label{introduction}}

Bayesian inferences are based upon summaries of a posterior distribution. Computing
this distribution and summaries of it requires evaluating a
difficult integral---called the normalizing constant or marginal
likelihood---and this computation is infeasible to do exactly in all but
the simplest of problems. Modern approaches for inference in a wide
range of models are based on approximations to the posterior. These
approximations, in turn, often rely on approximations to the normalizing
constant.

Quadrature refers to the approximation of an integral of a function (called the \emph{integrand}) using a 
finite weighted sum of evaluations of that function. A \emph{quadrature rule} is a set of \emph{nodes} at
which the function is to be evaluated and \emph{weights} by which the function
evaluations are to be multiplied when summing. There is a rich literature on 
quadrature rules and their properties from a numerical analysis perspective \citep{numint} 
as well as on their use in statistical 
problems \citep{evansintegrations}, and this literature may be drawn upon when choosing a rule to use
for integrating a chosen function. Of particular importance is ensuring that the 
quadrature rule places high weight on evaluations which are near to
where the function takes its largest values. This is necessary to obtain an accurate approximation to the integral.

Because the normalizing constant 
is an integral, in principle any quadrature
technique may be used to approximate it and (hence) make approximate Bayesian inferences. 
However, the posterior is defined by a particular model and prior and its
location and shape are random variables which vary with
the data. In practice, therefore, the region where the posterior takes its largest values will change with different
models, priors, and data. It follows that no fixed quadrature rule can be expected to work well for
all posterior distributions, even those obtained from the same model and prior but different datasets. 
Care is required when applying quadrature in the contex of Bayesian inference.

Adaptive Quadrature
techniques take any fixed quadrature rule and shift and scale its nodes and weights 
according to the mode and curvature of the integrand, attempting to automatically focus
attention on the region where the integrand takes its largest
values. The goal of adaptive quadrature is to obtain an accurate and useful approximation to a wider
variety of integrals than any single fixed quadrature rule can. Application of adaptive quadrature 
to the approximation of the normalizing constant is therefore a potentially
useful techinque for making approximate Bayesian inferences. 

Both the theoretical and
practical advantages of adaptive quadrature as a tool for making approximate Bayesian inferences
 have recently been studied by \citet{aghqus}. These authors show that adaptation
of a broad class of fixed quadrature rules using the posterior mode and curvature yields a method
for making approximate Bayesian inferences that has compelling asymptotic properties under
standard conditions on the model. They also motivate the use of adaptive quadrature 
in challenging Bayesian inference problems through several examples, which are considered in
greater computational detail in \S\ref{sec:examples} and \S\ref{sec:elgms} of the present
manuscript.

When the fixed quadrature rule used is Gauss-Hermite Quadrature (\GHQ{}),
 the corresponding adaptive rule is called Adaptive Gauss-Hermite Quadrature
(\texttt{AGHQ}{}). Since \texttt{AGHQ}{} was introduced in the
statistical literature
\citep{nayloradaptive,laplace,adaptive_GH_1994,adaptive_GH_2020} it has
been used in a variety of applications including latent variable models
\citep{aghqmle} and as the default option for approximating the likelihood
in Generalized Linear Mixed Models in the \pkg{glmer} function in the
popular \pkg{lme4} package \citep{lme4}. While
the theoretical analysis of \citet{aghqus} applies to any quadrature rule which satisfies certain
properties, their examples focus on \AGHQ{}. This choice is partly due to some 
desirable properties of \GHQ{} (\S\ref{sec:background}), the connection of \AGHQ{} with the Laplace
approximation, and the prevalence of \AGHQ{} in the existing literature.
This focus on \AGHQ{} is the case for the remainder
of this paper as well, although other quadrature rules including those based on sparse grids
and sampling-based evaluations could in principle replace \GHQ{} in all of
what follows and may be implemented in ongoing future releases of the \pkg{aghq} package.

Like all quadrature techniques, \GHQ{} and hence \texttt{AGHQ}{} are only directly
useful for models in which the dimension of the parameter space is not too large. 
However, one interesting property of \texttt{AGHQ}{} specifically is that
it recovers the \emph{Laplace approximation} when run
with only a single quadrature point, avoiding the unfavourable increase in computation
time that usually occurs with increasing dimension. When the dimension of the parameter space is
large, several applications of \AGHQ{} with different numbers of quadrature points can be applied,
recovering the \emph{marginal} Laplace approximation of \citet{laplace}. This approach is available in
the \pkg{aghq} package and is applied to high-dimensional \emph{latent Gaussian} models \citep{inla} in \S\ref{sec:loaloa} and 
\emph{extended latent Gaussian} models \citep{lgmsplit,maxandsmooth,noeps} in \S\ref{sec:loaloazip}. 
The \pkg{aghq} therefore provides users with a template for making approximate Bayesian
inferences in a wide range of challenging, high-dimensional models.

There are several existing packages which implement Adaptive Quadrature to
some extent, however they are all either too specific or too general for the present setting,
applying only to specific models or lacking functions for
using the results to make approximate Bayesian inferences. For example,
the \pkg{fastGHQuad} package \citep{fastghquad} implements
\texttt{AGHQ}{} for general scalar functions only, requires the user to
supply their own mode and curvature information, 
and does not include any
functions for post-processing the results. The \pkg{GLMMadaptive}
package \citep{glmmadaptive} handles optimization and has a formula
interface and an API for post-processing model results, but only fits
Generalized Linear Mixed Models with independent random effects. The
\pkg{LaplacesDemon} package provides an option to use \texttt{AGHQ}{} to
improve moment estimation in models fit either with \texttt{MCMC}{} or a
Laplace Approximation, but does not offer further use of the technique. 
Finally, the popular \pkg{INLA} package \citep{inla,inlasoftware} makes approximate
Bayesian inferences for a class of latent Gaussian additive models, implementing several types
of adaptive quadrature among a rich collection of approximations. The interface
is aimed at a general scientific audience and comparatively less of a focus is 
placed on specialists who wish to
implement their own models, although doing so is possible for models which are 
compatible with the underlying methodology.


In this paper the new \pkg{aghq} package is introduced. The \pkg{aghq} package contains an interface for
approximating the normalizing constant using \texttt{AGHQ}{}, and making approximate 
Bayesian inferences based on the result. For users who are implementing
their own priors and likelihoods, the \pkg{aghq} package
is simple to use, yet flexible enough to make approximate Bayesian inferences
in a wide range of problems not covered by existing software. Standard \texttt{print},
\texttt{summary}, and \texttt{plot} methods are provided for output objects of class \texttt{aghq}.
A flexible interface for taking \texttt{aghq} objects and
computing moments, quantiles, and marginal densities/distribution
functions is also provided. Further, in certain cases, fast independent samples from the approximate posterior are provided as
well, enabling the estimation of complicated posterior summaries not covered by other methods in the package.

While not a requirement, the \pkg{aghq} package has been designed to work especially
well in combination with the popular \pkg{TMB} package \citep{tmb} which
offers an interface for automatic differentiation and unnormalized marginal Laplace
approximations in \pkg{R}. Use of \pkg{TMB} handles the computation of the two derivatives
required to implement Adaptive Quadrature and which can otherwise be challenging to
obtain in the types of complicated models that \pkg{aghq} was designed to fit. It also
provides an automatic Laplace approximate marginal posterior and an efficient, 
exact gradient of this quantitity. Implementing approximate Bayesian inference for complicated additive
models (\S\ref{high-dimensions}) is straightforward using \pkg{TMB} in combination with \pkg{aghq}, as demonstrated in \S\ref{sec:elgms}. 
Users who already have their unnormalized log-posteriors coded in \pkg{TMB} can made approximate Bayesian inferences
of a form similar to \citet{inla} and \citet{noeps} using only two additional lines of code.

Further,
while unrelated to the \pkg{aghq} package, it is important to point out that use of
\pkg{TMB} to implement the unnormalized log-posterior further allows implementation of \MCMC{} for the same model
using no additional code with the \pkg{tmbstan} package \citep{tmbstan}, allowing comparison of \AGHQ{}
and \MCMC{} for problems in which the latter is computationally feasible. Therefore the use of \pkg{TMB}
embeds \pkg{aghq} within a comprehensive framework for performing the challenging
computations associated with making approximate Bayesian inferences. All of the examples
in this paper include comparisons to \MCMC{} and all but one are implemented in \pkg{TMB},
illustrating this approach.

The remainder of this paper is organized as follows. In \S\ref{sec:basic-use} the basic use of the \pkg{aghq} package is described in
detail using the simple example from the simulations in \citet{aghqus}.
In \S\ref{sec:background} the necessary background on
\texttt{AGHQ}{} is provided, including explicit formulas and a brief statement of some 
relevant mathematical properties. In \S\ref{sec:examples}
two examples from \citet{aghqus} are implemented using the \pkg{aghq} package, providing
users with a detailed template describing the use of the package for
making approximate Bayesian inferences. In \S\ref{sec:elgms}
the Generalized Linear Geostatistical Model (GLGM) considered by \citet{geostatsp} and \citet{prevmap} is implemented
 using \pkg{aghq}, and then extended to fit the much more challenging 
zero-inflated binomial 
geostatistical model described by \citet{geostatlowresource}. In
\S\ref{sec:glmmTMB} \pkg{aghq} is used to make approximate
Bayesian inferences in a situation where other authors \citep{glmmTMB} have already implemented the model in \pkg{TMB},
which is a main intended use case of the package. The paper concludes in
\S\ref{sec:discussion} with a discussion of impact and future work.

\hypertarget{sec:basic-use}{%
\section{Basic Use}\label{sec:basic-use}}

In this section the basic use of the \pkg{aghq} package is demonstrated by
way of analyzing a simple example in which the answer is known. Detailed discussion of
Bayesian inference and quadrature in \texttt{R} deferred to
\S\ref{sec:background}. All notation and terms are defined in detail in
that section.

The following conjugate model is used in simulations by \citet{aghqus}:
\begin{equation}\begin{aligned}
Y_i \ \vert \ \lambda &\overset{ind}{\sim} \text{Poisson}(\lambda), i\in[n], \\
\lambda &\sim \text{Exponential}(1), \lambda > 0, \\
\implies \lambda \ \vert \ \boldsymbol{Y}&\sim \text{Gamma}\left(1 + \sum_{i=1}^{n}y_{i},n+1\right). \\
\end{aligned}\end{equation}
The empirical accuracy of the method will depend on how close to log-quadratic the posterior distribution
is, with the method being more accurate for posteriors which are closer to being log-quadratic.
While the constraint that $\lambda > 0$ does not pose any barrier to application of \AGHQ{} in theory,
box constraints of this type are often obvious indicators that a parameter transformation may yield
a posterior that is closer to being log-quadratic, and hence for which \AGHQ{} will provide more accurate
results \citep{nayloradaptive}. Here we set $\theta = \log\lambda$, perform
the normalization on this scale, and then use the features of the \pkg{aghq}
package to make inferences for $\lambda = e^{\theta}$. It is important to reiterate that this step is 
not strictly \emph{necessary}, but is a good strategy to employ in practice.

The goal of the analysis is to approximate \[\pi(\boldsymbol{Y}) = \int_{0}^{\infty}\frac{\lambda^{\sum_{i=1}^{n}Y_{i}}e^{-(n+1)\lambda}}{\prod_{i=1}^{n}Y_{i}!}d\lambda = \frac{\Gamma\left( \sum_{i=1}^{n}Y_{i} + 1\right)}{(n+1)^{\sum_{i=1}^{n}Y_{i} + 1}\prod_{i=1}^{n}Y_{i}!},\] and then make Bayesian inferences
based on \(\pi(\lambda|\boldsymbol{Y})\) (\S\ref{sec:background}). First, install and load the
\pkg{aghq} package:

\begin{CodeChunk}
\begin{CodeInput}
R> # Install stable version from CRAN:
R> # install.packages("aghq")
R> # or development version from github:
R> # devtools::install_github("awstringer1/aghq")
R> library(aghq)
\end{CodeInput}
\end{CodeChunk}

Simulate some data from the true data generating process:

\begin{CodeChunk}
\begin{CodeInput}
R> set.seed(84343124)
R> y <- rpois(10,5) # True lambda = 5, n = 10
\end{CodeInput}
\end{CodeChunk}

The main function in the \pkg{aghq} package is \texttt{aghq::aghq()}. The user supplies
a list containing the log-posterior and its first two derivatives:

\begin{CodeChunk}
\begin{CodeInput}
R> logpithetay <- function(theta,y) {
+   sum(y) * theta - (length(y) + 1) * exp(theta) - sum(lgamma(y+1)) + theta
+ }
R> objfunc <- function(x) logpithetay(x,y)
R> objfuncgrad <- function(x) numDeriv::grad(objfunc,x)
R> objfunchess <- function(x) numDeriv::hessian(objfunc,x)
R> # Now create the list to pass to aghq()
R> funlist <- list(
+   fn = objfunc,
+   gr = objfuncgrad,
+   he = objfunchess
+ )
\end{CodeInput}
\end{CodeChunk}
as well as the number
of quadrature points, \(k\), and a starting value for the optimization.
Inference using \AGHQ{} is then performed:
\begin{CodeChunk}
\begin{CodeInput}
R> # AGHQ with k = 3
R> # Use theta = 0 as a starting value
R> thequadrature <- aghq::aghq(ff = funlist,k = 3,startingvalue = 0)
\end{CodeInput}
\end{CodeChunk}
The object \texttt{thequadrature} has class \texttt{aghq}, with
\texttt{summary} and \texttt{plot} methods:

\begin{CodeChunk}
\begin{CodeInput}
R> # Not shown
R> # summary(thequadrature)
R> # plot(thequadrature)
\end{CodeInput}

% \begin{center}\includegraphics{01-aghq-softwarepaper_files/figure-latex/doaghq2-1} \end{center}
% \begin{center}\includegraphics{01-aghq-softwarepaper_files/figure-latex/doaghq2-2} \end{center}

\end{CodeChunk}

The object \texttt{thequadrature} contains all quantities necessary for computation of
approximate moments of any function $g(\theta)$, approximate quantiles and marginal probability
densities and cumulative distribution functions of $\theta$ and any monotone transformation of it
(including for each parameter in multidimensional models), and exact samples from these approximate
marginals using the inverse CDF method/probability integral transform. Note that much of this information is displayed easily
using the \texttt{summary} and \texttt{plot} methods for objects of class \texttt{aghq}. 
How to obtain these quantities one at a time, using the interface for \texttt{aghq} objects
provided by the \pkg{aghq} package, is now described.

\begin{CodeChunk}
\begin{CodeInput}
R> # The normalized posterior at the adapted quadrature points:
R> thequadrature$normalized_posterior$nodesandweights
\end{CodeInput}
\begin{CodeOutput}
    theta1   weights   logpost logpost_normalized
1 1.246489 0.2674745 -23.67784         -0.3566038
2 1.493925 0.2387265 -22.29426          1.0269677
3 1.741361 0.2674745 -23.92603         -0.6047982
\end{CodeOutput}
\begin{CodeInput}
R> # The log normalization constant:
R> thequadrature$normalized_posterior$lognormconst
\end{CodeInput}
\begin{CodeOutput}
[1] -23.32123
\end{CodeOutput}
\begin{CodeInput}
R> # Compare to the truth: 
R> lgamma(1 + sum(y)) - (1 + sum(y)) * log(length(y) + 1) - sum(lgamma(y+1))
\end{CodeInput}
\begin{CodeOutput}
[1] -23.31954
\end{CodeOutput}
\begin{CodeInput}
R> # The mode found by the optimization:
R> thequadrature$optresults$mode
\end{CodeInput}
\begin{CodeOutput}
[1] 1.493925
\end{CodeOutput}
\begin{CodeInput}
R> # The true mode:
R> log((sum(y) + 1)/(length(y) + 1))
\end{CodeInput}
\begin{CodeOutput}
[1] 1.493925
\end{CodeOutput}
\end{CodeChunk}

Of course, in most problems, the true mode and normalizing constant
won't be known.

The \pkg{aghq} package provides further routines for computing moments,
quantiles, marginal probability densities and cumulative
distribution functions (including for monotone transformations), and exact independent samples from the approximate marginal posteriors. These routines are especially useful when
working with transformations like we are in the example, since interest
here is in the original parameter \(\lambda = e^{\theta}\). The
functions are all \texttt{S3} methods with a method for objects of class \texttt{aghq}. These are as follows:

\begin{itemize}
\item
  \texttt{compute\_pdf\_and\_cdf}: compute the density and cumulative
  distribution function for a marginal posterior distribution of $\theta$ 
  and (optionally) a smooth monotone transformation of that
  variable;
\item
  \texttt{compute\_moment}: compute the posterior moment of any
  function of $\theta$;
\item
  \texttt{compute\_quantiles}: compute posterior quantiles for a
  marginal posterior distribution.
\end{itemize}
For multidimensional parameters, all of these functions work on the associated multiple marginal posterior distributions, without any additional input from the user. 

To compute the approximate marginal
posterior for \(\lambda\),
\(\widetilde{\pi}_{\texttt{AGHQ}}(\lambda|\boldsymbol{Y})\), we first
compute the marginal posterior for \(\theta\) and then transform:
\begin{equation}
\widetilde{\pi}_{\texttt{AGHQ}}(\lambda|\boldsymbol{Y}) = \widetilde{\pi}_{\texttt{AGHQ}}(\theta|\boldsymbol{Y})\left\lvert \frac{\mathrm{d}\theta}{\mathrm{d}\lambda} \right\rvert.
\end{equation}

The \texttt{aghq::compute\_pdf\_and\_cdf()} function has an option,
\texttt{transformation}, which allows the user to specify a parameter
transformation that they would like the marginal density of. The user
specifies functions \texttt{fromtheta} and \texttt{totheta} which
convert from and to the transformed scale (on which quadrature was
done), and the function returns the marginal density on both scales,
making use of a numerically differentiated jacobian. This is all done as
follows:

\begin{CodeChunk}
\begin{CodeInput}
R> # Compute the pdf for theta
R> transformation <- list(totheta = log,fromtheta = exp)
R> pdfwithlambda <- compute_pdf_and_cdf(
+  thequadrature,
+  transformation = transformation
+ )[[1]]
R> lambdapostsamps <- 
+    sample_marginal(thequadrature,1e04,transformation = transformation)[[1]]
R> head(pdfwithlambda,n = 2)
\end{CodeInput}
\begin{CodeOutput}
      theta         pdf          cdf transparam pdf_transparam
1 0.9990534 0.008604132 0.000000e+00   2.715710    0.003168281
2 1.0000441 0.008809832 8.728201e-06   2.718402    0.003240813
\end{CodeOutput}
\begin{CodeInput}
R> # Plot the approximate density
R> # and samples along with the true posterior
R> with(pdfwithlambda,{
+   hist(lambdapostsamps,breaks = 50,freq = FALSE,main = "",xlab = expression(lambda))
+   lines(transparam,pdf_transparam)
+   lines(transparam,dgamma(transparam,1+sum(y),1+length(y)),lty='dashed')
+ })
\end{CodeInput}
\begin{figure}
\centering 
\includegraphics[width=0.45\textwidth]{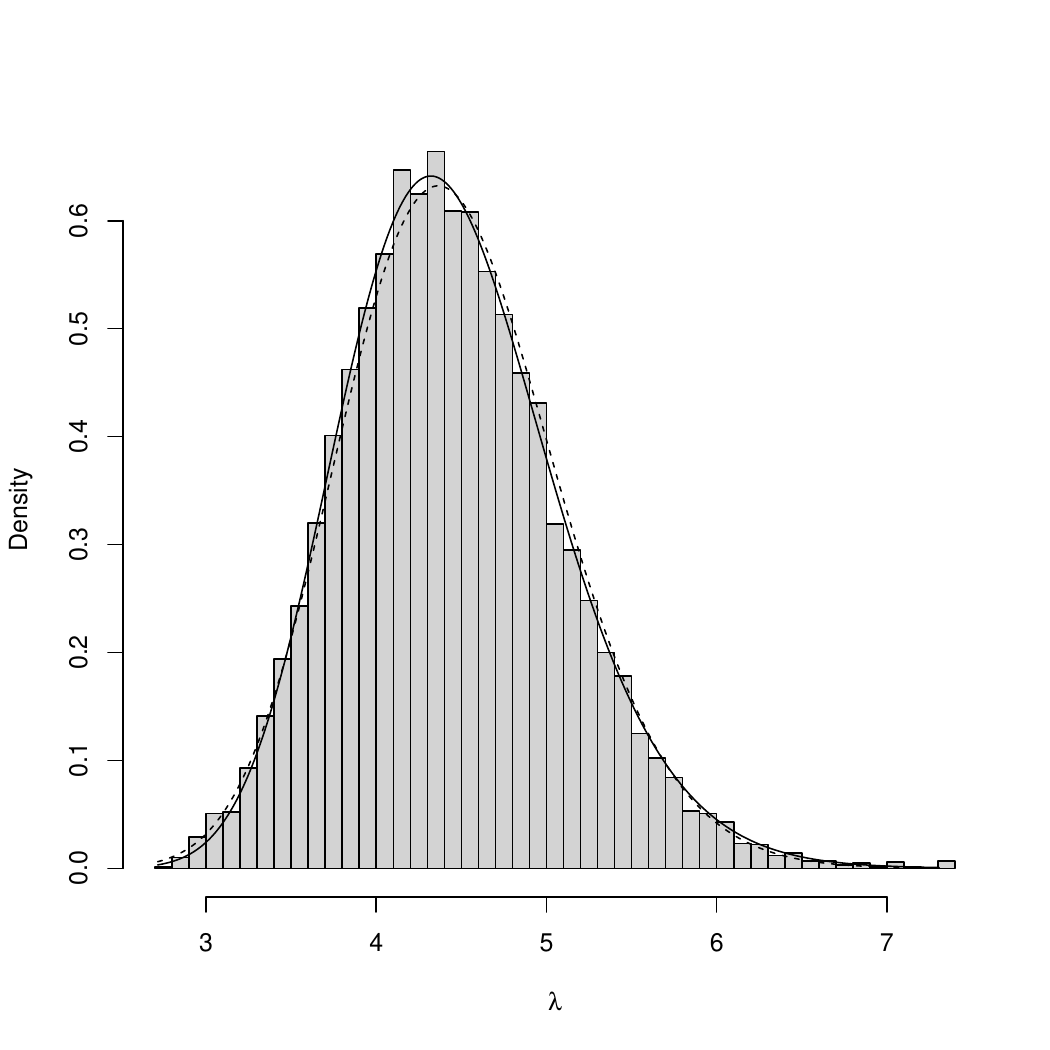}
\caption{Approximate (---, $\textcolor{lightgray}{\blacksquare}$) and true (- - -) posterior $\pi(\lambda|\mb{Y})$ in the example of \S\ref{sec:basic-use}.}
\label{fig:postforlambda1}
\end{figure}
\end{CodeChunk}
The need to index element \texttt{[[1]]} from the output
here feels messy, but is necessary to keep the interface
consistent between single- and multi-dimensional models.

We may compute the posterior mean of \(\lambda = e^{\theta}\),
\(\mathop{\mathrm{\mathbb{E}}}(\lambda|\boldsymbol{Y})\) by using the
\texttt{compute\_moment} function:

\begin{CodeChunk}
\begin{CodeInput}
R> # Posterior mean for lambda = exp(theta)
R> compute_moment(thequadrature$normalized_posterior,
+                ff = function(x) exp(x))
\end{CodeInput}
\begin{CodeOutput}
[1] 4.454407
\end{CodeOutput}
\begin{CodeInput}
R> # Compare to the truth:
R> (sum(y) + 1)/(length(y) + 1)
\end{CodeInput}
\begin{CodeOutput}
[1] 4.454545
\end{CodeOutput}
\end{CodeChunk}
The argument \texttt{ff} is passed through \texttt{match.fun()} so that \texttt{ff = exp} and \texttt{ff = 'exp'} are also both
valid input.

Quantiles are computed using the \texttt{compute\_quantiles()} function.
For example, to get the $1\%$ and $99\%$ percentiles as well as the median
and first and third quartiles of $\lambda$:

\begin{CodeChunk}
\begin{CodeInput}
R> # Quantiles
R> compute_quantiles(
+    thequadrature,
+    q = c(.01,.25,.50,.75,.99),
+    transformation = transformation
+  )[[1]]
\end{CodeInput}
\begin{CodeOutput}
      1%      25%      50%      75%      99% 
3.166469 4.000544 4.404081 4.848323 6.149735
\end{CodeOutput}
\begin{CodeInput}
R> # The truth:
R> qgamma(c(.01,.25,.50,.75,.99),1+sum(y),1+length(y))
\end{CodeInput}
\begin{CodeOutput}
[1] 3.108896 4.010430 4.424279 4.865683 6.067076
\end{CodeOutput}
\end{CodeChunk}

The estimation of quantiles, especially extreme quantiles, is much more
difficult than that of moments, and this is reflected in the small
differences in accuracy observed between the two in this example.

\hypertarget{sec:background}{%
\section{Background and Details}\label{sec:background}}

In this section, the underlying methodology of
\texttt{AGHQ}{} is reviewed with an emphasis on computational details.
The notation follows \citet{aghqus} who describe the methodological background in much greater detail. 

\hypertarget{bayesian-inference}{%
\subsection{Bayesian Inference}\label{bayesian-inference}}

Let \(\boldsymbol{\theta}\in\Theta \subseteq \mathbb{R}^{d}\) denote a
parameter, let \(\pi(\boldsymbol{\theta})\) denote its prior
density, and let \(\boldsymbol{Y}= \left\{ Y_{i}:i\in[n] \right\}\)
denote a sample with likelihood
\(\pi(\boldsymbol{Y}|\boldsymbol{\theta})\).
Bayesian inferences are made from summaries of the posterior distribution:
\begin{equation}\label{eqn:posterior}
\pi(\boldsymbol{\theta}|\boldsymbol{Y}) = \frac{\pi^{*}(\boldsymbol{\theta},\boldsymbol{Y})}{\int_{\Theta}\pi^{*}(\boldsymbol{\theta},\boldsymbol{Y})\mathrm{d}\boldsymbol{\theta}},
\end{equation} where
\(\pi^{*}(\boldsymbol{\theta},\boldsymbol{Y}) = \pi(\boldsymbol{Y}|\boldsymbol{\theta})\pi(\boldsymbol{\theta})\).
The denominator of (\ref{eqn:posterior}) is called the normalizing
constant, denoted by \(\pi(\boldsymbol{Y})\): \begin{equation}
\pi(\boldsymbol{Y}) = \int_{\Theta}\pi^{*}(\boldsymbol{\theta},\boldsymbol{Y})\mathrm{d}\boldsymbol{\theta},
\end{equation} and this is infeasible to compute exactly unless the model is very simple. 

Inferences are made from summaries of $\pi(\boldsymbol{\theta}|\boldsymbol{Y})$, which are themselves (intractable) integrals.
The posterior mean of any function $f:\Theta\to\R$ is
$$
\mathop{\mathrm{\mathbb{E}}}[f(\boldsymbol{\theta})|\boldsymbol{Y}] = \int_{\Theta}f(\mb{\theta})\pi(\boldsymbol{\theta}|\boldsymbol{Y})d\mb{\theta},
$$
and for any $t\in[d],\alpha\in[0,1]$ the marginal posterior $\alpha$-quantile of $\theta_{t}$ is the number $q_{t}^{\alpha}$ which satisfies
$$
\int_{\R^{d-1}\times(-\infty,q_{t}^{\alpha})}\pi(\boldsymbol{\theta}|\boldsymbol{Y})d\mb{\theta}_{-t}d\theta_{t} = \alpha.
$$
The posterior mean $\mathop{\mathrm{\mathbb{E}}}(\theta_{t}|\boldsymbol{Y})$ or median $q_{t}^{0.5}$ are common point estimates of 
the components $\theta_{t}$, and the posterior standard deviation or extreme quantiles are commonly used to quantify uncertainty. The \pkg{aghq} package provides a framework in which \(\pi(\boldsymbol{Y})\) is approximated
using \texttt{AGHQ}{} and approximate Bayesian inferences are made using such summary statistics 
calculated from the corresponding approximation to $\pi(\mb{\theta}|\mb{Y})$.

\hypertarget{subsec:quadrature}{%
\subsection{Quadrature}\label{subsec:quadrature}}

For any function \(f:\mathbb{R}^{d}\to\mathbb{R}\), a quadrature
approximation with $k\in\N$ points to its integral is given by the weighted sum:
\begin{equation}
\int_{A}f(\boldsymbol{x})\mathrm{d}\boldsymbol{x} \approx \sum_{\mb{x}\in\Qdk}f(\boldsymbol{x})\omega(\mb{x}),
\end{equation} where \(A\subseteq \mathbb{R}^{d}\) is any set,
\(\Qdk\subset\R^{d}\) is the set of \emph{nodes}, and
\(\omega:\Qdk\to\R\) are \emph{weights}. In
\pkg{R}, quadrature is performed using the flexible interface provided
by the \pkg{mvQuad} package, which computes the nodes and weights for different
quadrature rules with $d,k\in\N$. For example, with $d=1,k=3$, \GHQ{} is implemented as:

\begin{CodeChunk}
\begin{CodeInput}
R> # Integrate exp(-x^2) on (-Inf,Inf)
R> f <- function(x) exp(-(x^2)/2)
R> # Define a grid for Gauss-Hermite quadrature
R> # with k = 3 points
R> gg <- mvQuad::createNIGrid(1,"GHe",3)
R> # Perform the quadrature
R> mvQuad::quadrature(f,gg) 
\end{CodeInput}
\begin{CodeOutput}
[1] 2.51
\end{CodeOutput}
\begin{CodeInput}
R> # True value is sqrt(2pi)
R> sqrt(2*pi)
\end{CodeInput}
\begin{CodeOutput}
[1] 2.51
\end{CodeOutput}
\end{CodeChunk}

When $d > 1$, such univariate rules are extended to multiple dimensions,
and there are a number of ways in which this can be done. Here we focus
only on \emph{product rule} extensions, in which the nodes are repeated
in each dimension and the weights multiplied, defining $\Qdk = \Qk^{d}$ and
$\omega(\mb{x}) = \prod_{j=1}^{d}\omega(x_{j})$. Using this technique with
\(\abs{\Qk} = k\) points per dimension and a
\(d\)-dimensional integrand therefore requires \(\abs{\Qdk} = k^{d}\) function
evaluations, and this is computationally expensive. 
Methods based on sparse grids \citep{sparsegrids} or
employing more detailed computations to compute the nodes and weights \citep{designedquadrature} may
alleviate this challenge somewhat, however these are not yet 
implemented in the \pkg{aghq} package. The present approach to
inference in high-dimensional additive models is discussed in \S\ref{high-dimensions}
with examples in \S\ref{sec:elgms}.

\hypertarget{gauss-hermite-quadrature}{%
\subsection{Gauss-Hermite Quadrature}\label{gauss-hermite-quadrature}}

Gauss-Hermite Quadrature (\texttt{GHQ}{}; \citet{numint}) is a useful
quadrature rule based on the theory of polynomial interpolation. Define
the \emph{Hermite polynomials}
\[H_{k}(x) = (-1)^k e^{x^{2}/2}\frac{d^{k}}{dx^{k}}e^{-x^{2}/2}\] where
\(k\in\mathbb{N}\). The \texttt{GHQ}{} nodes and weights with \(k\)
quadrature points are: \begin{equation}\begin{aligned}
\Qk &= \left\{ x\in\mathbb{R}: H_{k}(x) = 0 \right\}, \\
\omega(x) &= \frac{k!}{H_{k+1}(x)^{2} \times \phi(x)}, j\in[k],
\end{aligned}\end{equation} where \(\phi(\cdot)\) is a standard normal
density. Some interesting properties of \texttt{GHQ}{} are as follows:
\citep{numint}:

\begin{enumerate}
\item $x = 0$ is a node if and only if $k$ is odd;
\item $-x$ is a node if and only $x$ is;
\item $\omega_{j} > 0$ for each $j\in[k]$;
\item for a polynomial $P_{\alpha}(x)$ of degree $\alpha \leq 2k-1$, 
$$\int_{\mathbb{R}}P_{\alpha}(x)\phi(x)\mathrm{d}x = \sum_{x\in\Qk}P_{\alpha}(x)\phi(x)\omega(x),$$
and no other quadrature rule with the same number of points $k$ can obtain this exactness property for polynomials of higher degree;
\item for any suitable function $f(\cdot)$, there exists some $\xi\in\mathbb{R}$ such that
$$\int_{\mathbb{R}}f(x)\phi(x)\mathrm{d}x \equiv \sum_{x\in\Qk}f(x)\phi(x)\omega(x) + f^{(2k)}(\xi)E_{k},$$
where $E_{k}$ is an error term depending on $k$ with $E_{k}\downarrow 0$ as $k\uparrow\infty$.
\end{enumerate}
The natural equivalents of several of these properties hold in multiple dimensions, 
although formally stating them requires more advanced notation and definitions.
See \citet{aghqus} for further detail.

\hypertarget{adaptive-gauss-hermite-quadrature}{%
\subsection{Adaptive Gauss-Hermite
Quadrature}\label{adaptive-gauss-hermite-quadrature}}

The unnormalized posterior \(\pi^{*}(\boldsymbol{\theta},\boldsymbol{Y})\)
has a location and curvature which vary stochastically with
\(\boldsymbol{Y}\), and no fixed quadrature rule will perform well for
all datasets and models. A rule which adapts to the changing shape and
location of \(\pi^{*}(\boldsymbol{\theta},\boldsymbol{Y})\) is
desirable. The Laplace approximation \citep{laplace} is possibly the most well-known
example of such a technique, in which $\pi^{*}(\boldsymbol{\theta},\boldsymbol{Y})$
is approximated using an appropriately shifted and scaled Gaussian distribution
and this is used to approximate $\pi(\mb{Y})$. This is closely related to 
a more general discussion of adaptive quadrature techniques which I now briefly describe.

Adaptive quadrature techniques shift and scale
the nodes and weights of a given fixed quadrature rule
according to the mode and curvature
of \(\pi^{*}(\boldsymbol{\theta},\boldsymbol{Y})\). Define
\begin{equation}\begin{aligned}
\widehat{\boldsymbol{\theta}}&= \text{argmax}_{\boldsymbol{\theta}}\log\pi^{*}(\boldsymbol{\theta},\boldsymbol{Y}), \\
\boldsymbol{H}&= -\partial^{2}_{\boldsymbol{\theta}}\log\pi^{*}(\widehat{\boldsymbol{\theta}},\boldsymbol{Y}), \\
\boldsymbol{H}^{-1} &= \boldsymbol{L}\boldsymbol{L}^{\tiny{\texttt{T}}},
\end{aligned}\end{equation} where \(\boldsymbol{L}\) is the (unique)
lower Cholesky triangle. For a given rule with nodes $\Qdk$ and weights $\omega$, the corresponding
adaptive quadrature approximation to $\pi(\mb{Y})$ is:
\begin{equation}
\widetilde{\pi}(\mb{Y}) = \abs{\mb{L}}\sum_{\mb{x}\in\Qdk}\pi(\mb{Lx} + \widehat{\mb{\theta}},\mb{Y})\omega(\mb{x}).
\end{equation}
When the fixed quadrature rule is \GHQ{}, the method is called 
Adaptive Gauss-Hermite Quadrature (\texttt{AGHQ}{}; \citet{nayloradaptive,adaptive_GH_1994,adaptive_GH_2020,aghqus}), and 
the corresponding approximation is denoted $\widetilde{\pi}_{\AGHQ{}}(\mb{Y})$. 
This is the focus of the remainder of this paper. Some interesting properties of \texttt{AGHQ}{} are as follows (although
they may be shared by other adaptive rules as well):

\begin{enumerate}
\item
  The mode \(\widehat{\boldsymbol{\theta}}\) is a quadrature point
  if and only if \(k\) is odd;
\item
  When \(k = 1\), \texttt{AGHQ}{} is a Laplace approximation, providing a principled link between
  adaptive quadrature in low and high dimensions (\S\ref{high-dimensions}, \S\ref{sec:elgms});
\item
  When \(k > 1\), \texttt{AGHQ}{} achieves higher-order asymptotic
  accuracy: \citet{aghqus} show that if
  \(\widetilde{\pi}_{\texttt{AGHQ}}(\boldsymbol{Y})\) is an
  \texttt{AGHQ}{} approximation to \(\pi(\boldsymbol{Y})\) using \(k\)
  points, then \begin{equation}
  \widetilde{\pi}_{\texttt{AGHQ}}(\boldsymbol{Y}) = \pi(\boldsymbol{Y})\left[1 + \mathcal{O}_{p}\left( n^{-\lfloor (k+2)/3\rfloor}\right) \right],
  \end{equation} where the convergence is in probability measured with
  respect to the distribution of \(\boldsymbol{Y}\). Table
  \ref{tab:ratetable} shows the asymptotic accuracy achieved for
  selected \(k\). Note that the well-known \(n^{-1}\) rate for the
  Laplace approximation is recovered when \(k = 1\). Note that their 
  analysis applies to any rule that satisfies property 4.
  from \S\ref{gauss-hermite-quadrature} (polynomial exactness), not only \GHQ{}.
\end{enumerate}

\begin{table}
\centering
\begin{tabular}{|l|r|}
\hline
$k$ & Rate \\
\hline
1 & $\mathcal{O}_{p}\left(n^{-1}\right)$ \\
3 & $\mathcal{O}_{p}\left(n^{-1}\right)$ \\
5 & $\mathcal{O}_{p}\left(n^{-2}\right)$ \\
7 & $\mathcal{O}_{p}\left(n^{-3}\right)$ \\
9 & $\mathcal{O}_{p}\left(n^{-3}\right)$ \\
11 & $\mathcal{O}_{p}\left(n^{-4}\right)$ \\
13 & $\mathcal{O}_{p}\left(n^{-5}\right)$ \\
15 & $\mathcal{O}_{p}\left(n^{-5}\right)$ \\
\hline
\end{tabular}
\caption{Asymptotic rates for \texttt{AGHQ}{}}
\label{tab:ratetable}
\end{table}

\hypertarget{posterior-summaries}{%
\subsection{Posterior summaries}\label{posterior-summaries}}

With the \texttt{AGHQ}{} approximation
\(\widetilde{\pi}_{\texttt{AGHQ}}(\boldsymbol{Y})\) to
\(\pi(\boldsymbol{Y})\), the approximate posterior density is
\begin{equation}
\widetilde{\pi}_{\texttt{AGHQ}}(\boldsymbol{\theta}|\boldsymbol{Y}) = \frac{\pi^{*}(\boldsymbol{\theta},\boldsymbol{Y})}{\widetilde{\pi}_{\texttt{AGHQ}}(\boldsymbol{Y})},
\end{equation} and, by definition, satisfies \begin{equation}
\abs{\mb{L}}\sum_{\mb{x}\in\Qdk}\widetilde{\pi}_{\texttt{AGHQ}}(\mb{Lx} + \widehat{\boldsymbol{\theta}}|\boldsymbol{Y})\omega(\mb{x}) = 1.
\end{equation} For any function \(f:\mathbb{R}^{d}\to\mathbb{R}\), the
posterior moment
\(\mathop{\mathrm{\mathbb{E}}}[f(\boldsymbol{\theta})|\boldsymbol{Y}]\)
may be approximated by 
\begin{equation}
\mathop{\mathrm{\mathbb{E}}}[f(\boldsymbol{\theta})|\boldsymbol{Y}]\approx \abs{\mb{L}}\sum_{\mb{x}\in\Qdk}f(\mb{Lx} + \widehat{\boldsymbol{\theta}})\widetilde{\pi}_{\texttt{AGHQ}}(\mb{Lx} + \widehat{\boldsymbol{\theta}}|\boldsymbol{Y})\omega(\mb{x}),
\end{equation}
and this is what the \texttt{compute\_moment()} function does (\S\ref{sec:basic-use}). 
Note that this is not an application of \AGHQ{} to the integral defining this expectation; \citet{laplace} (among others)
discuss approximating such summaries which occur as ratios of integrals, however approximations of this type are not
implemented in the \pkg{aghq} package. The re-use of the \AGHQ{} points and weights in computation of moments
is not yet covered by theory, but appears empirically accurate in applications (\S\ref{sec:examples}).

Marginal posteriors and quantiles are more difficult to compute, as described in
detail by \citet{nayloradaptive} and \citet{aghqus}. The challenge is that the 
adaptive quadrature
rule is only defined at specific points of the form
\(\mb{Lx} + \widehat{\mb{\theta}}, \mb{x}\in\Qdk\), and these points are not on a
regular grid. It is
therefore only straightforward to evaluate marginal posteriors at \(k\)
distinct points, and this only works for the first element of
\(\boldsymbol{\theta}\). To mitigate these challenges, the parameter
vector has to be re-ordered and the joint normalization re-computed for
each desired marginal distribution, although the optimization results are reused. 
A polynomial or spline-based interpolant to 
these points is then used for plotting and computing quantiles. See \citet{aghqus}
for further detail. As described in \S\ref{sec:basic-use}, the
\pkg{aghq} package handles these computations for the user through the
\texttt{compute\_pdf\_and\_cdf()} and \texttt{compute\_quantiles()} functions.

\hypertarget{high-dimensions}{%
\subsection{High-dimensional additive models}\label{high-dimensions}}

A very important use case and motivation for development of the \pkg{aghq} package is the fitting
high-dimensional additive models by combining the Laplace ($k=1$) and \AGHQ{} ($k>1$) approximations.
This is inspired by the popular \texttt{INLA} approach of \citet{inla} for latent Gaussian models, 
and further explored by 
\citet{casecrossover,noeps} for the broader class of extended latent Gaussian models introduced by \citet{lgmsplit,maxandsmooth}.
The applications of \citet{casecrossover,noeps} include semi-parametric regression with multinomial response, analysis of
aggregated spatial point process data, survival analysis with spatially-varying hazard, and an astrophysical measurement error model for
estimating the mass of the Milky Way. Other examples
include non-Gaussian model-based geostatistics with zero-inflated observations \citep{aghqus}, and
large-scale spatio-temporal models \citep{inlamra} where the comparison
to \texttt{INLA} is made more explicitly. Here the core approach to inference using \pkg{aghq} is described, with details
to follow for the individual models fit in \S\ref{sec:elgms}.

Consider a model with parameter $(\mb{W},\mb{\theta})\in\R^{m}\times\R^{d}$ where $m\gg d$. It is
assumed that $d$ is small enough to reasonably apply \AGHQ{} with $k>1$ to integrals involving $\mb{\theta}$ (\S\ref{sec:basic-use}, \S\ref{sec:background}),
but that $m$ is so large as to render this computationally infeasible for integrals involving $\mb{W}$, even if using sparse grids
or other multidimensional extensions more efficient than the product rule we use here. The posteriors of interest are:
\begin{equation}\begin{aligned}
\pi(\mb{\theta}|\mb{Y}) &\propto \int_{\R^{m}}\pi(\mb{W},\mb{\theta},\mb{Y})d\mb{W}, \\
\pi(\mb{W}|\mb{Y}) &= \int_{R^{d}}\pi(\mb{W}|\mb{\theta},\mb{Y})\pi(\mb{\theta}|\mb{Y})d\mb{\theta}.
\end{aligned}\end{equation}
Applying \AGHQ{} with $k=1$ to $\pi(\mb{\theta}|\mb{Y})$ yields the (unnormalized) \emph{marginal Laplace approximation}
of \citet{laplace}, $\widetilde{\pi}_{\LA}(\mb{\theta},\mb{Y})$, and
then normalizing this using \AGHQ{} with $k>1$ gives:
\[
\widetilde{\pi}_{\LA}(\mb{\theta}|\mb{Y}) = \frac{\widetilde{\pi}_{\LA}(\mb{\theta},\mb{Y})}{\abs{\mb{L}}\sum_{\mb{x}\in\Qdk}\widetilde{\pi}_{\LA}(\mb{Lx} + \widehat{\mb{\theta}},\mb{Y})\omega(\mb{x})}.
\]
A suitable Gaussian approximation is used for $\pi(\mb{W}|\mb{\theta},\mb{Y}) \approx \widetilde{\pi}_{\G}(\mb{W}|\mb{\theta},\mb{Y})$.
The \AGHQ{} nodes and weights are reused in the final approximation:
\[\label{eqn:finalapproxelgm}
\pi(\mb{W}|\mb{Y})\approx\widetilde{\pi}(\mb{W}|\mb{Y}) = \abs{\mb{L}}\sum_{\mb{x}\in\Qdk}\widetilde{\pi}_{\G}(\mb{W}|\mb{Lx}+\widehat{\mb{\theta}},\mb{Y})\widetilde{\pi}_{\LA}(\mb{Lx}+\widehat{\mb{\theta}}|\mb{Y})\omega(\mb{x}).
\]
Inferences for $\mb{W}$ and any function of it are made by drawing exact posterior samples from the mixture-of-Gaussians approximation $\widetilde{\pi}(\mb{W}|\mb{Y})$. The full algorithm is given by \citet{noeps}. The INLA approach of \citet{inla} then involves approximating the marginal posteriors $\pi(W_{j}|\mb{Y}),j\in[m]$ using a further Laplace approximation, but this is not yet implemented in \pkg{aghq}.

The \pkg{aghq} package contains an interface for performing these challenging computations. The normalized marginal Laplace approximation $\widetilde{\pi}_{\LA}(\mb{\theta}|\mb{Y})$ is obtained using \texttt{aghq::marginal\_laplace()}. This function internally defines $\log\widetilde{\pi}_{\LA}(\mb{\theta},\mb{Y})$ using
\texttt{aghq::laplace\_approximation()} and then calls \texttt{aghq::aghq()}. However, it uses numerical derivatives of $\log\widetilde{\pi}_{\LA}(\mb{\theta},\mb{Y})$, and hence requires repeated optimizations of $\log\pi(\mb{W},\mb{\theta},\mb{Y})$ at great computational expense. The use of \texttt{marginal\_laplace()} is recommended for simpler models in which this strategy is feasible.

The recommended general approach to implementing Equation (\ref{eqn:finalapproxelgm}) is to implement $-\log\pi(\mb{W},\mb{\theta},\mb{Y})$ as a function template in the popular \pkg{TMB} package \citep{tmb} and turn on the \texttt{random} flag for $\mb{W}$. This provides $-\log\widetilde{\pi}_{\LA}(\mb{\theta},\mb{Y})$ and its exact gradient automatically, avoiding repeated expensive optimizations over $\mb{W}$. The \texttt{aghq::marginal\_laplace\_tmb()} function accepts such
a template from \pkg{TMB} and utilizes these features, leading to more efficient computations for normalizing $\log\widetilde{\pi}_{\LA}(\mb{\theta},\mb{Y})$ using \AGHQ{}. An added benefit of this feature of the \pkg{aghq} package is that the many users who are already using \pkg{TMB} to make approximate Bayesian inferences using its built in Gaussian approximations can begin making more accurate approximate Bayesian inferences using \AGHQ{} with almost no additional code.

Both \texttt{aghq::marginal\_laplace()} and \texttt{aghq::marginal\_laplace\_tmb()} return objects inheriting from classes \texttt{aghq} and \texttt{marginallaplace}.
Inferences for $\mb{\theta}$ are therefore made using all the same functions as described in \S\ref{sec:basic-use}, which automatically work on $\widetilde{\pi}_{\LA}(\mb{\theta}|\mb{Y})$. Sample-based inferences for $\mb{W}$ are made using the \texttt{aghq::sample\_marginal()} function which acts on \texttt{marginallaplace} objects and draws fast, exact samples from $\widetilde{\pi}(\mb{W}|\mb{Y})$ by making use of quantities previously computed and saved by \texttt{aghq::marginal\_laplace*}. These samples are then used to estimate any posterior summary of interest. The entire approach is best illustrated through the examples of \S\ref{sec:elgms} and \S\ref{sec:glmmTMB}, which despite their complexity all require remarkably little code to implement beyond the likelihood and priors themselves with to the \pkg{aghq} package.

\hypertarget{sec:examples}{%
\section{Examples, low dimensions}\label{sec:examples}}

In this section the two low-dimensional examples considered by \citet{aghqus} are implemented using \pkg{aghq}.
The use of \pkg{aghq} is highlighted and results are compared to the Gaussian approximation
returned by \pkg{TMB}, as well as \MCMC{} through \pkg{tmbstan}. 
All code for these examples is available from \href{https://github.com/awstringer1/aghq-software-paper-code}{https://github.com/awstringer1/aghq-software-paper-code}; only brief illustrative code is included in the text here.

\hypertarget{sec:inf-disease}{%
\subsection{Infectious disease models}\label{sec:inf-disease}}

\citet{epi} implement \texttt{MCMC}{} for a
\emph{Susceptible, Infectious, Removed} (SIR) model for the spread of
infectious disease in the \pkg{EpiILMCT} package. Here
we illustrate how to use the \pkg{aghq} package to fit this model as done by \citet{aghqus}. 
We compare to 
\MCMC{} using the \pkg{tmbstan} package, finding that \AGHQ{} produces results
much faster than \MCMC{} with little loss in accuracy for this simple model.

We consider their example of an outbreak of the Tomato Spotted Wilt Virus in
\(n = 520\) plants, of which \(n_{0} = 327\) were infected. For observed
infection times $\mb{I} = \left\{ I_{i}:i\in[n]\right\}$ with \(I_{1} \leq \cdots \leq I_{n_{0}}\) and
\(I_{i} = \infty, i = n_{0}+1,\ldots,n\), as well as removal times
\(\mb{R} = \left\{R_{i},i\in[n]\right\}\), the likelihood is \begin{equation}
\pi(\boldsymbol{I},\boldsymbol{R} | \alpha,\beta) = \prod_{j=2}^{n_{0}}\left( \sum_{i:I_{i} < I_{j} \leq R_{i}}\lambda_{ij}\right)\exp\left\{ -\sum_{i=1}^{n_{0}}\sum_{j=1}^{n}[\text{min}(R_{i},I_{j}) - \text{min}(I_{i},I_{j})]\lambda_{ij} \right\},
\end{equation} where \(\lambda_{ij} = \alpha d_{ij}^{-\beta}\) with
\(d_{ij}\) the Euclidean distance between plants \(i\) and \(j\).
Independent \(\text{Exponential}(0.01)\) priors are placed on the
parameters of interest \((\alpha,\beta)\), and quadrature is performed
on the transformed scale with
\(\boldsymbol{\theta}= (\log\alpha,\log\beta)\). See \citet{aghqus} and
\citet{epi} for further information.

To implement \texttt{AGHQ}{} for this problem using the \pkg{aghq}
package, we define a list of functions \texttt{ff} which compute the log-posterior 
and its first two derivatives, and then call the \texttt{aghq} function 
as in \S\ref{sec:basic-use}:

\begin{CodeChunk}
\begin{CodeInput}
R> library(TMB)
R> data("tswv", package = "EpiILMCT")
R> # Create the functions
R> ff <- TMB::MakeADFun(...)
R> # Do the quadrature
R> quadmod <- aghq(ff,9,c(0,0),control = default_control(negate = TRUE))
\end{CodeInput}
\end{CodeChunk}

The template returned by \texttt{TMB::MakeADFun()} returns the \emph{negated}
unnormalized log-posterior $-\log\pi(\mb{\theta},\mb{Y})$, which is
required to use \pkg{tmbstan}. The \texttt{control = default\_control(negate = TRUE)}
option tells \texttt{aghq()} to account for this internally.

We can compute marginal moments, quantiles, and densities of $\alpha$ and $\beta$:

\begin{CodeChunk}
\begin{CodeInput}
R> # Mean of alpha and beta
R> compute_moment(quadmod,function(x) exp(x))
\end{CodeInput}
\begin{CodeOutput}
[1] 0.01203082 1.30371767
\end{CodeOutput}
\begin{CodeInput}
R> # Quantiles of alpha and beta
R> posttrans <- list(totheta = log,fromtheta = exp)
R> compute_quantiles(quadmod,transformation = posttrans)
\end{CodeInput}
\begin{CodeOutput}
[[1]]
       2.5%       97.5% 
0.007570042 0.016617720 

[[2]]
    2.5%    97.5% 
0.981385 1.582341
\end{CodeOutput}
\begin{CodeInput}
R> # Marginal densities
R> quaddens <- compute_pdf_and_cdf(quadmod,posttrans)
\end{CodeInput}
\end{CodeChunk}
Joint posterior moments may also be approximated: for example, for two
plants which are \(d_{ij} = 2\) units apart, the infectivity rate is
\(\lambda_{ij} = \alpha\cdot2^{-\beta}\). We can approximate
\(\mathop{\mathrm{\mathbb{E}}}(\lambda_{ij}|\boldsymbol{Y})\) as:
\begin{CodeChunk}
\begin{CodeInput}
R> compute_moment(
+   quadmod,
+   function(x) exp(x)[1] * 2^(-exp(x)[2])
+ )
\end{CodeInput}
\begin{CodeOutput}
[1] 0.004804631
\end{CodeOutput}
\end{CodeChunk}
The length of the return value of \texttt{compute\_moment()} equals the length of its second argument, the function
whose moment is being approximated. The value of each component is the approximate posterior moment of that component.

Finally, marginal posterior samples of size $M$ are obtained from $\widetilde{\pi}_{\AGHQ{}}(\alpha|\mb{Y})$ and $\widetilde{\pi}_{\AGHQ{}}(\beta|\mb{Y})$
using the univariate inverse CDF/probability integral transform:
\begin{CodeChunk}
\begin{CodeInput}
R> quadsamps <- sample_marginal(quadmod,M)
\end{CodeInput}
\end{CodeChunk}
These could be used to estimate any marginal posterior summary of interest.

\begin{figure}[t]
\centering
\subfloat[$\alpha$]{\includegraphics[width=.45\textwidth]{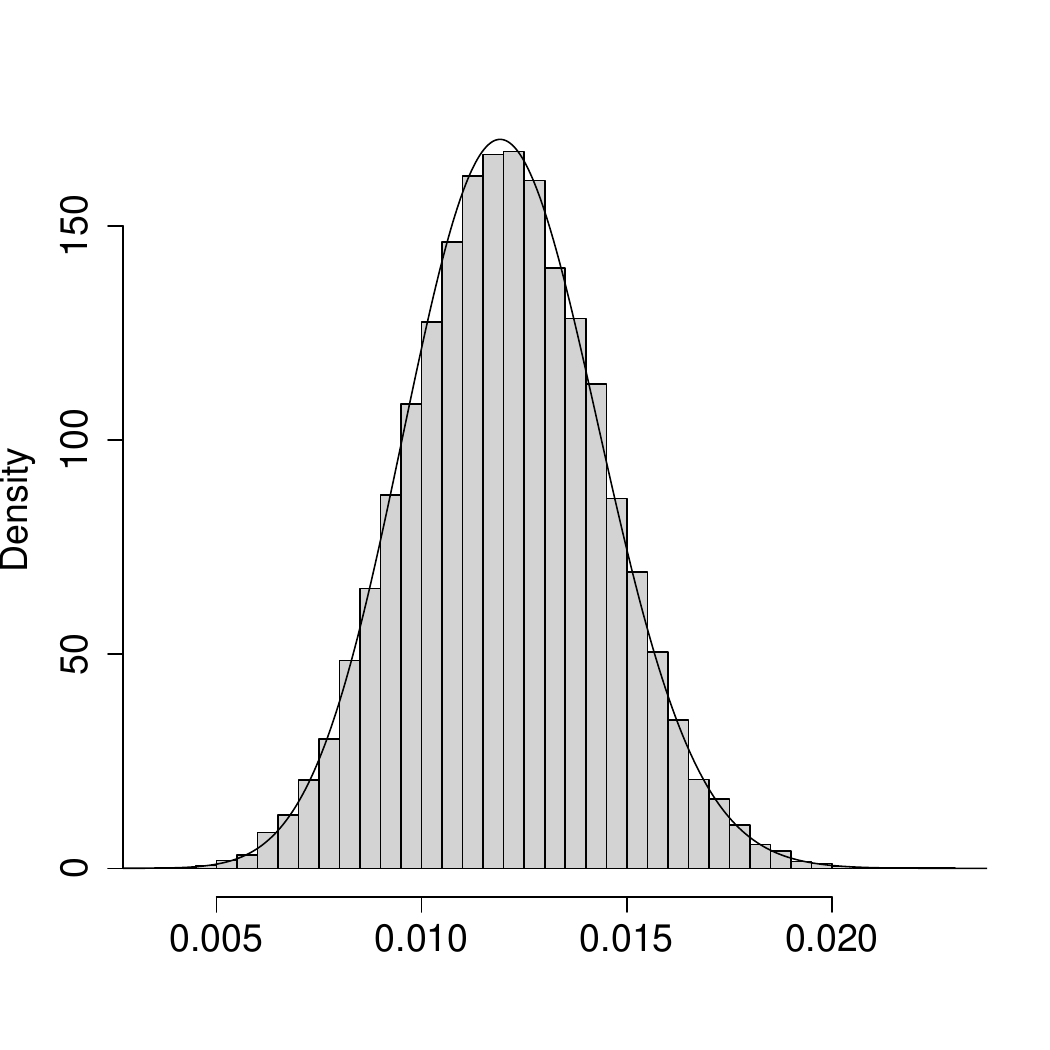}}
\subfloat[$\beta$]{\includegraphics[width=.45\textwidth]{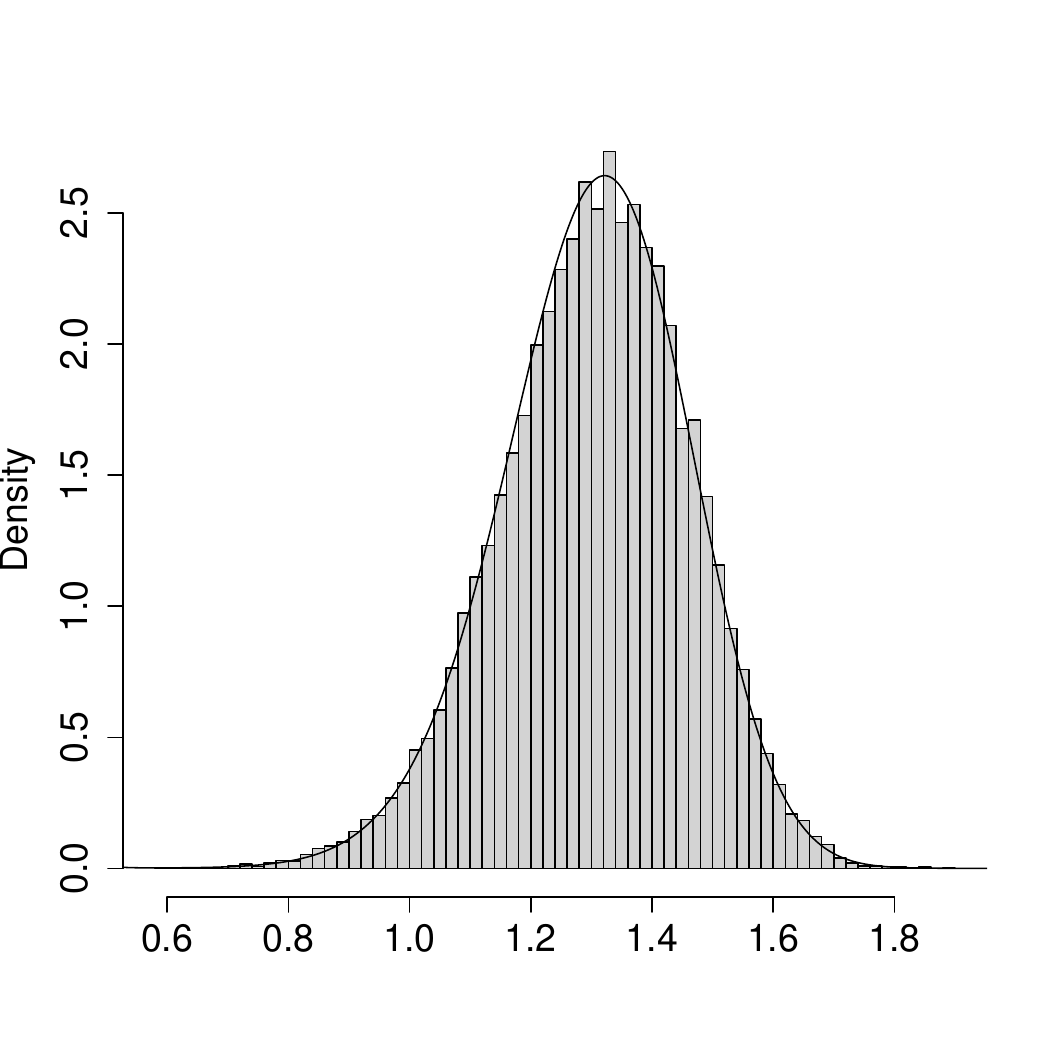}}
\caption{Approximate posteriors using \AGHQ{} (---) and \MCMC{} ($\textcolor{lightgray}{\blacksquare}$) for the infectious disease model example of \S\ref{sec:inf-disease}.}
\label{fig:diseasepostplots}
\end{figure}

Figure \ref{fig:diseasepostplots} shows the \AGHQ{} approximate marginal posteriors along with those obtained using \pkg{tmbstan}.
Table \ref{tab:diseasepostsumms} shows various summary statistics obtained using both methods, including the KS statistics obtained by
running \texttt{ks.test()} on the \MCMC{} and \AGHQ{} marginal posterior samples for both variables. The 
results based on \AGHQ{} are very close to those obtained using \MCMC{}. The runtime for \AGHQ{} was
$0.59$ seconds, compared to 330.88 seconds using \MCMC{} with four parallel chains of 10,000 iterations each (including warmup)
and all the default settings. Since the number of iterations
must be chosen and this changes the runtime of \MCMC{}, we base our conclusions about relative runtimes 
off of the observation that the time it took to fit the model
using \AGHQ{} is sufficient to run only $0.59\times 10000/330.88 = 17.84$ iterations of \MCMC{} using the chosen settings.
It appears that \AGHQ{} results in faster inferences with little loss of accuracy in this example.

\begin{table}[t]
\centering
\begin{tabular}{|l|rrrr|}
\hline
 & \multicolumn{2}{c}{$\alpha$} & \multicolumn{2}{c|}{$\beta$} \\
 & \AGHQ{} & \MCMC{} & \AGHQ{} & \MCMC{} \\
\hline
$\EE(\cdot|\mb{Y})$ & 0.0120 & 0.0121 & 1.30 & 1.31 \\
$\text{SD}(\cdot|\mb{Y})$ & 0.00232 & 0.00235 & 0.153 & 0.155 \\
$\widetilde{q}^{0.025}$ & 0.00757 & 0.0755 & 0.981 & 0.982 \\
$\widetilde{q}^{0.975}$ & 0.0166 & 0.0167 & 1.58 & 1.59 \\
$\EE(\alpha\cdot2^{-\beta}|\mb{Y})$ & 0.00480 & 0.00481 & - & - \\
\hline
KS(\AGHQ{},\MCMC{}) & \multicolumn{2}{r}{0.0195} & \multicolumn{2}{r|}{0.0235} \\
\hline
\end{tabular}
\caption{Comparison of approximate posterior summary statistics obtained using \AGHQ{} and \MCMC{} for the infectious disease model example of \S\ref{sec:inf-disease}.}
\label{tab:diseasepostsumms}
\end{table}

\hypertarget{subsec:astro}{%
\subsection{Estimating the mass of the Milky Way
Galaxy}\label{subsec:astro}}

Astrophysicists are interested in estimating the mass of the Milky Way
Galaxy. \citet{gwen2} describe a statistical model which relates
multivariate position/velocity measurements of star clusters orbiting
the Galaxy to parameters which in turn determine its mass according to a
physical model. Their interest is in assessing the
impact of various prior choices and data inclusion rules on inferences,
and they fit and assess many models using \MCMC{}. \citet{aghqus} implement
the model using \AGHQ{}. Here, the use of the \pkg{aghq} package to fit this model is demonstrated. The high accuracy 
attained (in theory) by \AGHQ{} is shown to produce results that are much closer to the
\MCMC{} samples returned by \pkg{tmbstan} than a simpler Gaussian approximation to the posterior used in \pkg{TMB}. Further, the \pkg{aghq} package gives these more accurate results using
about the same amount of code and with similar runtime to \pkg{TMB}. 

In this example the optimization required to implement \AGHQ{} is more challenging than in the example of \S\ref{sec:inf-disease}. The \pkg{aghq}
package accomodates more difficult optimization problems by allowing
the user to perform the optimization outside and pass in the results, and this feature is illustrated here. Some current limitations of the methods underpinning the \pkg{aghq} package are also illustrated
through the estimation and quantification of uncertainty for a complicated
nonlinear, transdimensional posterior summary using \AGHQ{} and \MCMC{}.

Let \(\boldsymbol{Z}_{i} = (z_{i1},\ldots,z_{i4})\) denote the position
and three-component velocity measured relative to the sun
(\emph{Heliocentric} measurements) for the \(i^{th}\) star cluster,
\(i\in[n], n = 70\). These are what is actually measured. These
measurements are then transformed into position and radial and
tangential velocity measurements relative to the centre of the Galaxy
(\emph{Galactocentric} measurements)
\(\boldsymbol{Y}_{i} = (y_{i1},y_{i2},y_{i3})\). The probability density
for \(\boldsymbol{Y}_{i}\) is \begin{equation}
f(\boldsymbol{Y}_{i};\mb{\Xi}) = \frac{L_{i}^{-2\beta}\mathcal{E}_{i}^{\frac{\beta(\gamma-2)}{\gamma} + \frac{\alpha}{\gamma} - \frac{3}{2}}\Gamma\left( \frac{\alpha}{\gamma} - \frac{2\beta}{\gamma} + 1\right)} {\sqrt{8\pi^{3}2^{-2\beta}} \Psi_{0}^{-\frac{2\beta}{\gamma} + \frac{\alpha}{\gamma}}\Gamma\left( \frac{\beta(\gamma - 2)}{\gamma} + \frac{\alpha}{\gamma} - \frac{1}{2}\right)},
\end{equation} where \(L_{i} = y_{i1}y_{i3}\) and
\(\mathcal{E}_{i} = \Psi_{0}y_{i1}^{1-\gamma} - (y_{i2}^{2} + y_{i3}^{2})/2\).
The parameters \(\boldsymbol{\Xi} = (\Psi_{0},\gamma,\alpha,\beta)\)
determine the mass of the Galaxy at radial distance \(r\) kiloparsecs
(kpc) from Galaxy centre as: \begin{equation}
M_{\boldsymbol{\Xi}}(r) = \Psi_{0}\gamma r^{1-\gamma},
\end{equation} according to a physical model, and inference is done for
\(M_{\boldsymbol{\Xi}}(r)\) at various values of \(r\).

\begin{figure}[t]
\centering
\makebox{
\subfloat[$\Psi_{0}$]{\includegraphics[width=.45\textwidth]{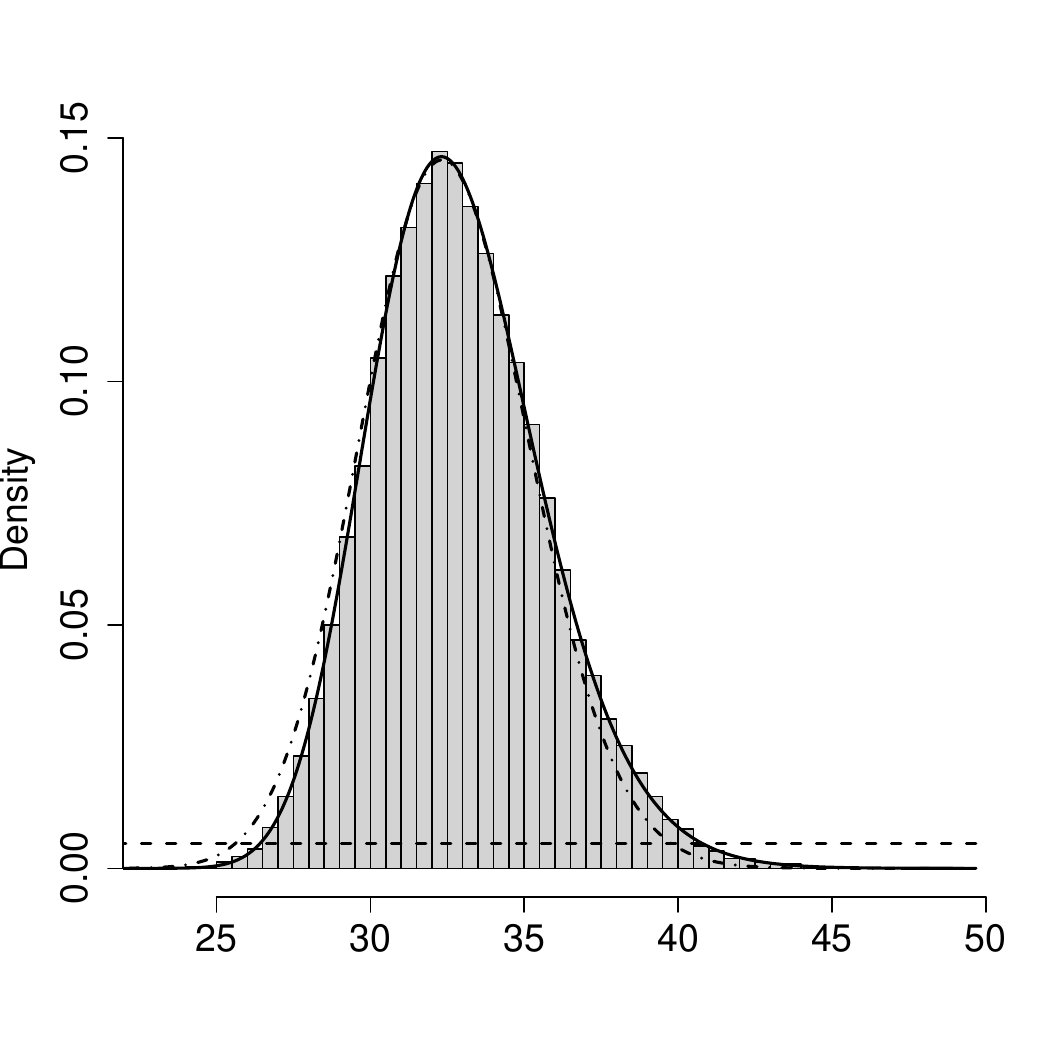}}
\subfloat[$\gamma$]{\includegraphics[width=.45\textwidth]{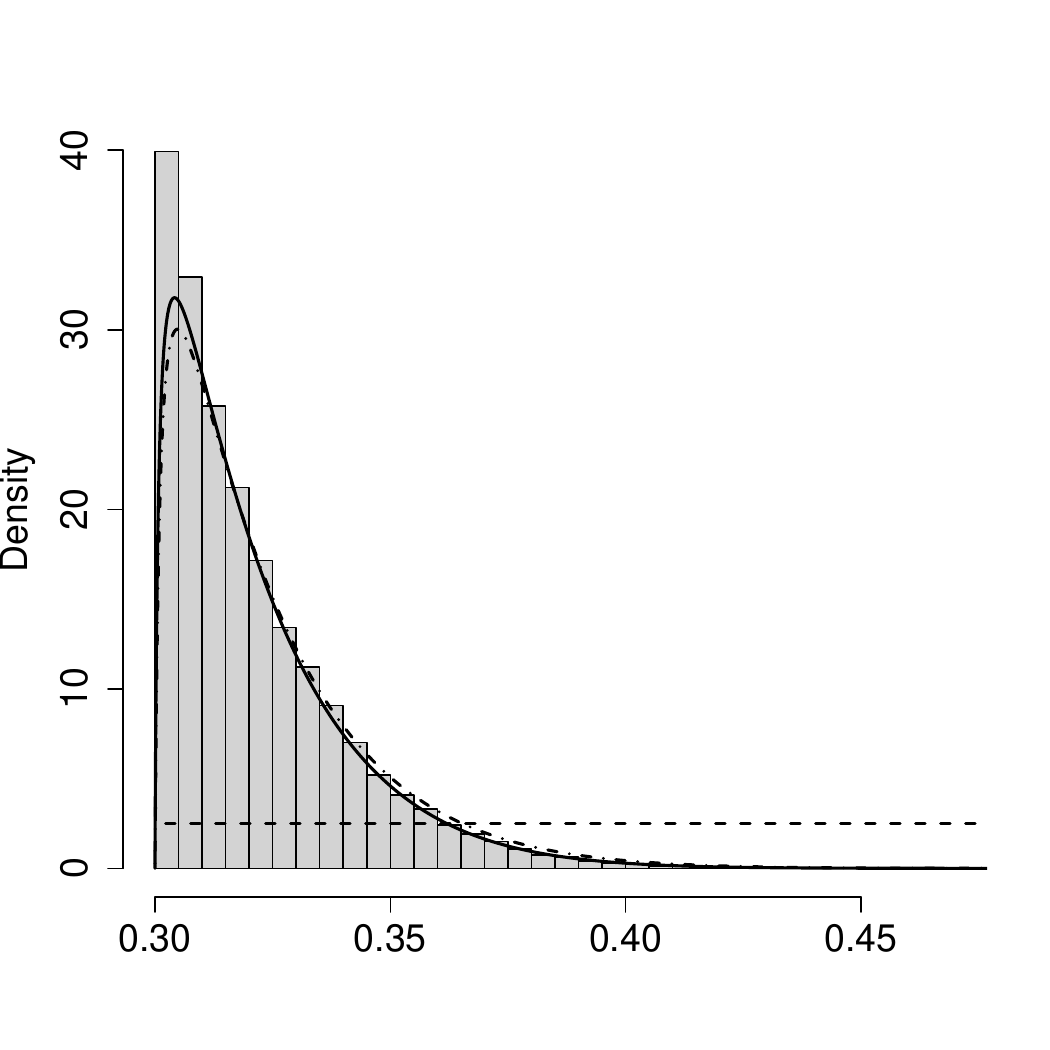}}
} \\
\makebox{
\subfloat[$\alpha$]{\includegraphics[width=.45\textwidth]{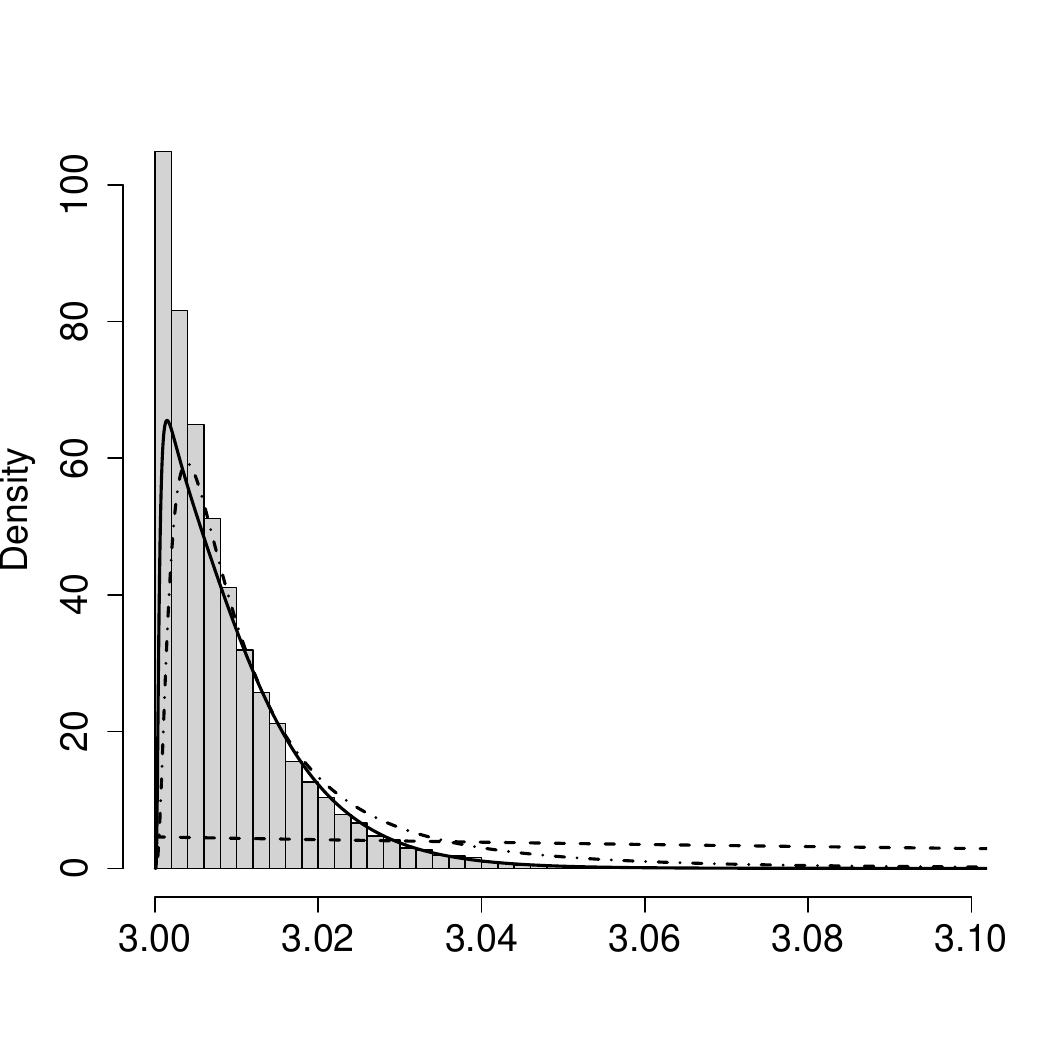}}
\subfloat[$\beta$]{\includegraphics[width=.45\textwidth]{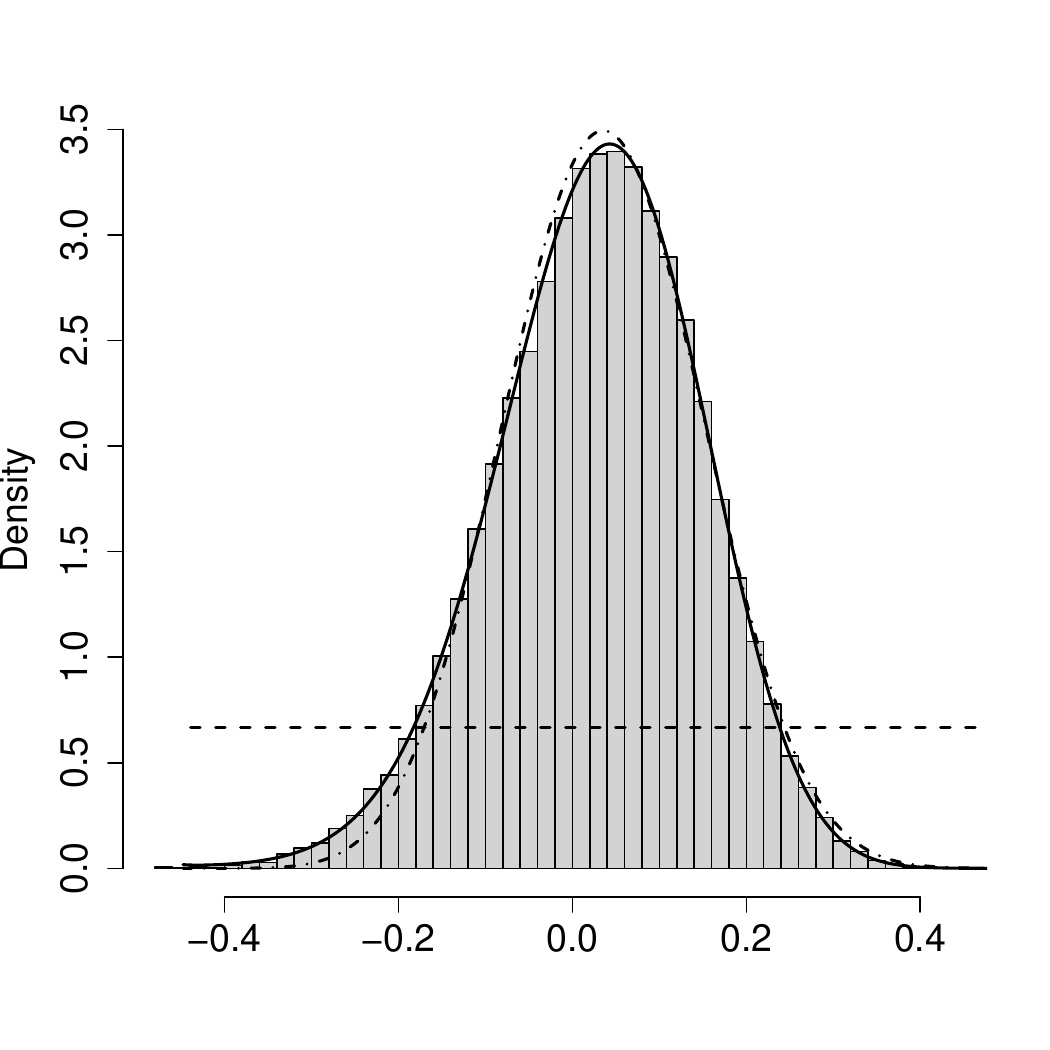}}
} \\
\caption{Approximate posteriors using \AGHQ{} (---), \TMB{} (-$\cdot$-) and \MCMC{} ($\textcolor{lightgray}{\blacksquare}$) for the Galaxy mass estimation example of \S\ref{subsec:astro}.}
\label{fig:astropostplots}
\end{figure}

The parameters are subject to nonlinear and box constraints:
\begin{equation}
\begin{aligned}
\alpha &>\gamma, \qquad \alpha > \beta(2 - \gamma) + \frac{\gamma}{2}, \qquad \mathcal{E}_{i} > 0, i\in[n], \\ 
\alpha &\geq 3, \qquad 1 \leq \Psi \leq 200, \qquad 0.3 \leq \gamma \leq 0.7, \qquad -0.5 \leq \beta \leq 1. \\
\end{aligned}
\end{equation} 
Uniform priors on these ranges are used for
\(\Psi,\gamma,\beta\), and a \(\text{Gamma}(1,4.6)\) prior is used for
\(\alpha - 3\). We perform quadrature on the transformed scale
\(\boldsymbol{\theta}= (\theta_{1},\theta_{2},\theta_{3},\theta_{4})\)
with \begin{equation}\begin{aligned}
\theta_{1} = \log\left(-\log\left[ \frac{\Psi - 1}{200 - 1}\right]\right), &\qquad \theta_{2} = \log\left(-\log\left[ \frac{\gamma - 0.3}{0.7 - 0.3}\right]\right), \\
\theta_{3} = \log(\alpha - 3), &\qquad \theta_{4} = \log\left(-\log\left[ \frac{\beta + 0.5}{1 + 0.5}\right]\right). \\
\end{aligned}\end{equation} This transformation enhances the spherical
symmetry of the posterior and we find it to improve the stability of the
optimization and the empirical quality of the \texttt{AGHQ}{}
approximation, as suggested by \citet{nayloradaptive}. The nonlinear
constraints remain, however, and none of the built-in optimizers provided
by \pkg{aghq} handle this. For this reason, optimization in this example is
performed manually using the much more substantial optimization interface 
from the \pkg{ipoptr} package \citep{ipopt}, and these results are
passed to \texttt{aghq::aghq()}. This works as follows:

\begin{CodeChunk}
\begin{CodeInput}
R> library(TMB)
R> data("gcdatalist",package = "aghq")
R> # Negated un-normalized log-posterior and its derivatives
R> ff <- TMB::MakeADFun(...,ADreport = FALSE)
R> # Nonlinear constraints and their jacobian
R> Es <- TMB::MakeADFun(...,ADreport = TRUE)
R> # Optimization
R> ipopt_result <- ipoptr::ipoptr(...)
R> useropt <- list(
+    ff = list(
+      fn = function(theta) -1*ff$fn(theta),
+      gr = function(theta) -1*ff$gr(theta),
+      he = function(theta) -1*ff$he(theta)
+    ),
+    mode = ipopt_result$solution,
+    hessian = ff$he(ipopt_result$solution)
+  )
R> # Quadrature
R> astroquad <- aghq(
+   ff,5,thetastart,optresults = useropt,control = default_control(negate=TRUE)
+  )
\end{CodeInput}
\end{CodeChunk}
A Gaussian approximation $\widetilde{\pi}_{\G}(\mb{\theta}|\mb{Y})$ uses a mean equal
to $\widehat{\mb{\theta}}$ and covariance matrix equal to $\mb{H}^{-1}$, with 
approximate marginal posteriors obtained from the corresponding marginal Gaussian
distributions. The information obtained from \pkg{TMB}:
\begin{CodeChunk}
\begin{CodeInput}
R> tmbsd <- TMB::sdreport(ff)
\end{CodeInput}
\end{CodeChunk}
is also present in the \pkg{aghq} output:
\begin{CodeChunk}
\begin{CodeInput}
R> # Matches tmbsd$par.fixed
R> astroquad$optresults$mode
R> # Matches sqrt(diag(tmbsd$cov.fixed))
R> sqrt(diag(solve(astroquad$optresults$hessian)))
\end{CodeInput}
\end{CodeChunk}
Figure \ref{fig:astropostplots} shows the approximate marginal posterior distributions of the four model parameters,
obtained using \MCMC{}, \texttt{TMB} and \AGHQ{}, 
and Table \ref{tab:astroks} shows the estimated KS distance between the \MCMC{} results
and samples from the approximate posteriors using \AGHQ{} and \TMB{}. In general, \AGHQ{}
is much closer to \MCMC{} than the Gaussian approximation; this is especially evident for $\alpha$,
which has a sharply skewed posterior and for which the Gaussian approximation is very inaccurate, especially
in the tails, where this would have an effect on calculation of credible intervals.
The runtime of \AGHQ{} was 1.40 seconds, taking 1.02 seconds for optimization using \pkg{ipopt}
and 0.38 seconds for quadrature; \TMB{} required no additional time beyond the optimization. The
\MCMC{} was run under the same conditions as Example \ref{sec:inf-disease} and had a runtime of 
21.9 seconds, meaning that \AGHQ{} could be run in the time it takes to run 640 \MCMC{} iterations,
while \TMB{} takes the same time as 468 iterations.

\begin{table}[t]
\centering
\begin{tabular}{|l|rr|}
\hline
Param. & \AGHQ{} & \TMB{} \\
\hline
$\Psi_{0}$ & 0.010 & 0.044 \\
$\gamma$ & 0.0073 & 0.031 \\
$\alpha$ & 0.035 & 0.16 \\
$\beta$ & 0.0099 & 0.025 \\
\hline
\end{tabular}
\caption{Estimated KS distances between the \MCMC{} results and those from \AGHQ{} and \TMB{} for the Galaxy mass estimation example of \S\ref{subsec:astro}.}
\label{tab:astroks}
\end{table}

The interest in this example is in the cumulative mass profile $M_{\boldsymbol{\Xi}}(r)$ at
a variety of values of distance from the centre of the Galaxy, $r$, in kiloparsecs (kPc). Estimates of the posterior mean
and standard deviation of this quantity are easily obtained using \texttt{aghq::compute\_moment()}:
\begin{CodeChunk}
\begin{CodeInput}
R> # Cumulative mass profile
R> Mr <- function(r,theta) {
+    p = get_psi0(theta[1])
+    g = get_gamma(theta[2])
+    # Manual unit conversion into "mass of one trillion suns"
+    g*p*r^(1-g) * 2.325e09 * 1e-12
+  }
R> # Posterior mean mass at 100 kPc from centre, for example:
R> compute_moment(astroquad,function(x) Mr(100,x))
\end{CodeInput}
\begin{CodeOutput}
[1] 0.5592737
\end{CodeOutput}
\end{CodeChunk}
Because $M_{\boldsymbol{\Xi}}(r)$ is a nonlinear, transdimensional summary of $\mb{\theta}$,
neither approximate credible intervals nor sample-based inference are available options
in the current release of \pkg{aghq}. The best we can currently do with the \pkg{aghq}
is estimate its mean and standard deviation and use the latter to quantify uncertainty in 
the former, which effectively amounts to assuming a Gaussian posterior for $M_{\boldsymbol{\Xi}}(r)$.
To investigate the quality of this approximation, posterior means and empirical
$2.5\%$ and $97.5\%$ quantiles were obtained from the \MCMC{} run as well. 

\begin{figure}[t]
\centering
\subfloat[\AGHQ{}]{\includegraphics[width=.45\textwidth]{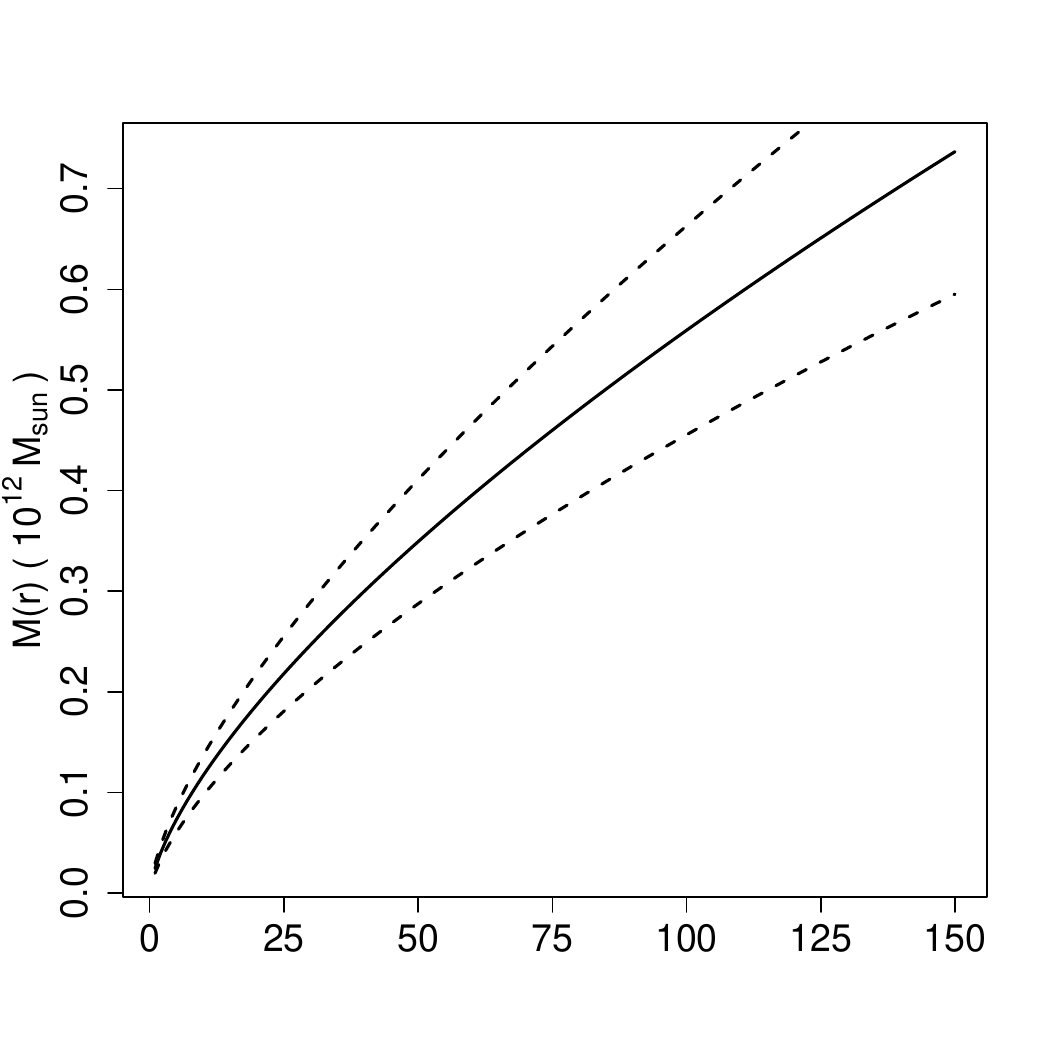}}
\subfloat[\MCMC{}]{\includegraphics[width=.45\textwidth]{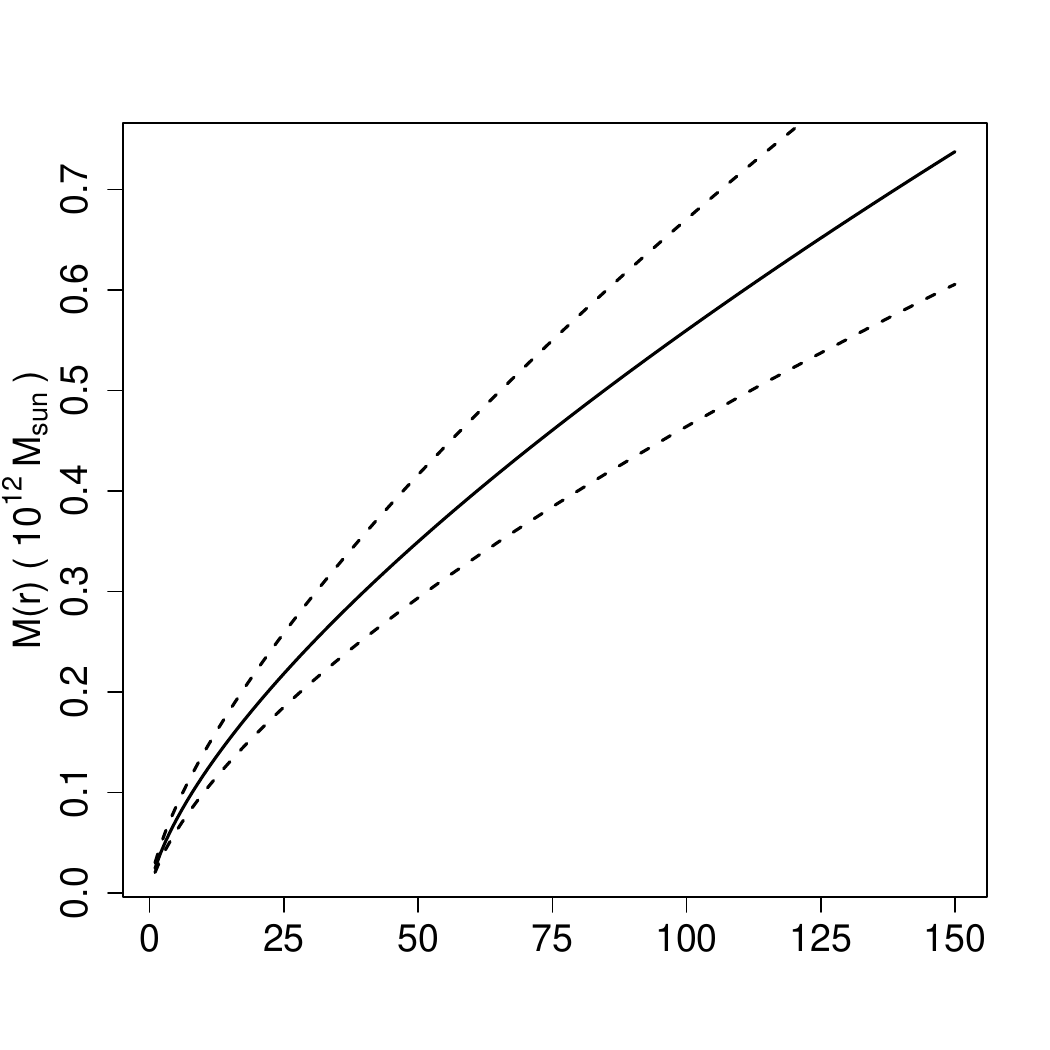}}
\caption{Estimated mass $M_{\boldsymbol{\Xi}}(r)$ of the Milky Way as a function of radial distance $r$ (kPc) from its centre (---) with approximate $95\%$ credible intervals (- - -)
using (a) \AGHQ{} and (b) \MCMC{} for the Galaxy mass estimation example of \S\ref{subsec:astro}. 
Credible intervals are computed using the estimate plus/minus 
twice the posterior standard deviation for \AGHQ{}, and by using sample $2.5\%$ and $97.5\%$ quantiles 
for \MCMC{}.}
\label{fig:massplots}
\end{figure}

Figure \ref{fig:massplots} shows
the two estimated mass profiles. While there is little apparent visual difference, note that the units of measurement
are ``the mass of one trillion suns'', so small visual differences may be consequential. 
The emprical root-mean square difference between the two types of estimated posterior means and lower/upper
credible intervals were 0.000412, 0.00743, and 0.0061 respectively, in these units. 
The present example illustrates a current limitation in the types
of inferences we can make using the \pkg{aghq} package. Note that these are \emph{methodological} limitations
rather than merely limitations with the current implementation. As more flexible methods are developed, 
they will be implemented in future releases of \pkg{aghq}.

\hypertarget{sec:elgms}{%
\section{Examples, high dimensions}\label{sec:elgms}}

While there are some models in which
\texttt{AGHQ}{} can provide the sole framework for Bayesian inference (\S\ref{sec:examples}), the \pkg{aghq} package is also useful for making approximate Bayesian inferences in certain
high-dimensional additive models (\S\ref{high-dimensions}). In \S\ref{sec:loaloa}, approximate Bayesian inferences for a \emph{latent Gaussian model} for disease mapping
with a non-Gaussian response are made using the \pkg{aghq} package. Comparisons are made to a similar model
fit using the \INLA{} method for approximate Bayesian inference \citep{inla} 
through the \texttt{geostatsp} package \citep{geostatsp}, as well as using maximum likelihood and \MCMC{} through the \texttt{PrevMap} package \citep{prevmap}.

However, 
the \pkg{aghq} package is significantly more flexible than existing packages in the breadth of models for which approximate Bayesian inferences can be made. This is demonstrated in \S\ref{sec:loaloazip} with the fitting of a
\emph{zero-inflated} version of the model from \S\ref{sec:loaloa}, originally introduced by \citet{geostatlowresource}
and later fit by \citet{aghqus}. This model is an \emph{extended latent Gaussian model} \citep{lgmsplit,maxandsmooth,noeps} and is challenging to fit. It is not compatible with \INLA{}, and
\MCMC{} can also be very challenging here; \citet{aghqus} discuss how off the shelf
\MCMC{} using \pkg{tmbstan} leads to divergent results after nearly three days of running. Using the interface provided in the \pkg{aghq} package (\S\ref{high-dimensions})
the model is fit in approximately 90 seconds using almost the same code as required to fit the much simpler,
non-zero inflated model.

\hypertarget{sec:loaloa}{%
\subsection{Geostatistical Binomial regression: the Loa Loa parasitic
roundworm in Cameroon and Nigeria}\label{sec:loaloa}}

Several authors \citep{loaloa,geostatsp,prevmap} have considered a dataset containing 
counts of subjects who tested positive for a tropical
disease caused by the \emph{Loa loa} parasitic roundworm in \(n = 190\)
villages in Cameroon and Nigeria. The goal is to model the spatial variation in
the probability of infection. For village $i\in[n]$ let $0\leq Y_{i}\leq N_{i}\in\N$ denote the number 
of infected and total number of residents, and $\mb{s}_{i}\in\R^{2}$ denote its spatial location.
Let $p(\mb{s})$ denote the probability of infection at location $\mb{s}\in\R^{2}$. The model is:
\begin{equation}\begin{aligned}
Y_{i} | p(\mb{s}_{i}) &\overset{ind}{\sim}\text{Binomial}[N_{i},p(\mb{s}_{i})], \\
\log\left[\frac{p(\mb{s})}{1-p(\mb{s})}\right] &= \beta_{0} + u(\boldsymbol{s}), \mb{s}\in\R^{2}, \\
u(\cdot) | \boldsymbol{\theta} &\sim \mathcal{GP}(0,\text{C}_{\boldsymbol{\Xi}}). \\
\end{aligned}\end{equation} 
The unknown process \(u(\cdot)\)
governs excess spatial variation in logit infection risk, and is
modelled as a Gaussian Process with Matern
covariance function \(\text{Cov}[u(\mb{s}+\mb{h}),u(\mb{s})] = \text{C}_{\boldsymbol{\Xi}}(\mb{h}), \mb{s},\mb{h}\in\R^{2}\) 
depending on parameters \(\boldsymbol{\Xi} = (\sigma,\rho)\) which
represent the marginal standard deviation and practical correlation range; see \citet{geostatsp}.
Quadrature is performed on the
transformed variable $\boldsymbol{\theta}= (\log\kappa,\log\tau)$ using the transformations suggested by
\citet{pcpriormatern}: 
\begin{equation}
\kappa = \sqrt{8\nu}/\rho, \qquad \tau = \sigma\kappa^{\nu} \sqrt{\frac{\Gamma\left(\nu + d/2\right)(4\pi)^{d/2}}{\Gamma(\nu)}}, \\
\end{equation} 
where \(\nu = 1, d = 2\) in this example.
Exponential priors for $\sigma$ and $\rho^{-1}$ are used following \citet{pcpriormatern} and satisfy
\(P(\sigma < 4) = P(\rho < 200\text{km}) = 97.5\%\), and a Gaussian prior is used for $\beta_{0}$.

The parameter of interest is $\mb{W} = (\mb{U},\beta_{0})$ in the notation of \S\ref{high-dimensions}, where
$\mb{U} = \bracevec{u(\mb{s}_{i}):i\in[n]}$. The \pkg{aghq} package is used to fit this model using the 
approximations defined in \S\ref{high-dimensions}. Functions computing
\(\log\pi(\boldsymbol{W},\boldsymbol{\theta},\boldsymbol{Y}), 
\partial_{\mb{W}}\log\pi(\boldsymbol{W},\boldsymbol{\theta},\boldsymbol{Y})\),
and $\partial^{2}_{\mb{W}}\log\pi(\boldsymbol{W},\boldsymbol{\theta},\boldsymbol{Y})$ are
defined with the following signature:

\begin{CodeChunk}
\begin{CodeInput}
R> ff <- list(
+   fn = function(W,theta) ...,
+   gr = function(W,theta) ...,
+   he = function(W,theta) ...
+ )
\end{CodeInput}
\end{CodeChunk}

and this list is then passed to the \texttt{aghq::marginal\_laplace()} function:

\begin{CodeChunk}
\begin{CodeInput}
R> # Starting values taken from Brown (2011)
R> startingsig <- .988
R> startingrho <- 4.22*1e04
R> startingtheta <- c(
+    log(get_kappa(startingsig,startingrho)),
+    log(get_tau(startingsig,startingrho))
R> )
R> Wstart <- rep(0,Wdim)
R> loaloaquad <- marginal_laplace(ff,3,list(W = Wstart,theta = thetastart))
\end{CodeInput}
\end{CodeChunk}
The \texttt{marginal\_laplace} function returns an object inheriting from class
\texttt{aghq} as usual, and inferences for \(\boldsymbol{\theta}\) may be made
 using the usual \texttt{summary} and \texttt{plot} methods. However,
interest in this example is for \(\boldsymbol{W}\). The returned object also inherits from class \texttt{marginallaplace} and contains additional information required to
draw samples from $\widetilde{\pi}(\mb{W}|\mb{Y})$ using the appropriate method dispatched by \texttt{aghq::sample\_marginal()}. Specifically, calling \texttt{aghq::sample\_marginal()} on an object inheriting from class \texttt{marginallaplace} efficienty draws a specified
number of samples from
\(\widetilde{\pi}(\boldsymbol{W}|\boldsymbol{Y})\):
\begin{CodeChunk}
\begin{CodeInput}
R> samps <- sample_marginal(loaloaquad,1e02)
\end{CodeInput}
\end{CodeChunk}
and returns a list containing these samples as well as those obtained from the marginals of $\widetilde{\pi}(\mb{\theta}|\mb{Y})$ as is usually done for \texttt{aghq} objects. The \emph{joint} posterior samples from \(\widetilde{\pi}(\boldsymbol{W}|\boldsymbol{Y})\) can then be post-processed on a problem-specific basis to estimate any
posterior functional of interest. In this example they are used as an input to
the \texttt{geostatsp::RFsimulate()} function which performs spatial
interpolation via conditional simulation; see \citet{aghqus}, \citet{geostatsp}, or the 
\texttt{RandomFields} package \citep{randomfields}.

Several similar methods are available for comparison in this example. There are minor differences
between each implementation although the results are broadly comparable. The \texttt{geostatsp} package \citep{geostatsp} provides an
interface to the \pkg{INLA} software which is used to implement the \INLA{}
method \citep{inla} for this model. This software
uses a basis function approximation to $u(\cdot)$ \citep{spde} instead of exact
point locations $u(\mb{s}_{i})$. The \texttt{PrevMap} package \citep{prevmap} implements
two additional methods: \emph{Monte Carlo Maximum Likelihood} (\MCML{}) and \MCMC{}, 
which fit similar models to what we fit here using 
\AGHQ{}.

\begin{figure}[t]
\centering
\makebox{
\subfloat[\AGHQ{}]{\includegraphics[width=.45\textwidth]{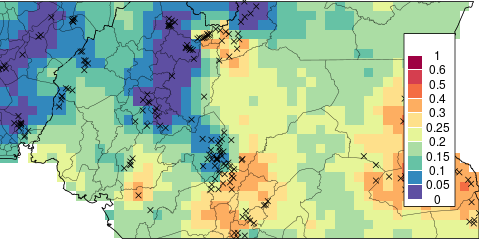}}
\subfloat[\MCMC{}]{\includegraphics[width=.45\textwidth]{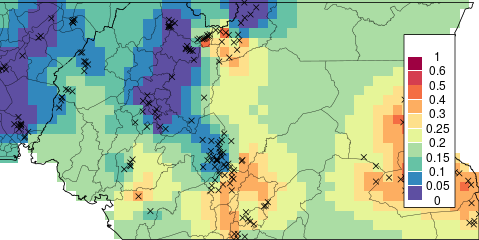}}
} \\
\makebox{
\subfloat[\MCML{}]{\includegraphics[width=.45\textwidth]{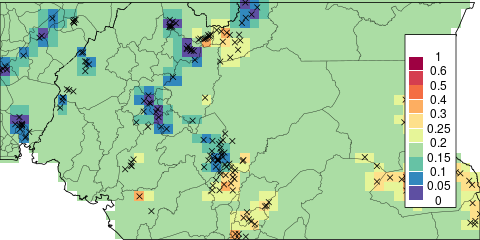}}
\subfloat[\INLA{}]{\includegraphics[width=.45\textwidth]{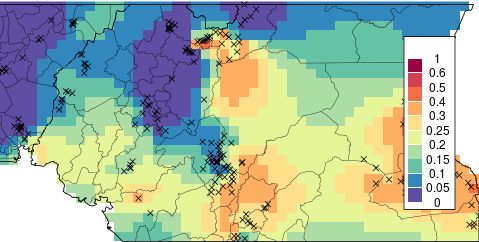}}
} \\
\caption{Spatial interpolations for the four different appraches to inference for the \emph{Loa loa} example
of \S\ref{sec:loaloa}.}
\label{fig:loaloamaps}
\end{figure}

Figure \ref{fig:loaloamaps} shows the resulting spatial interpolations of $p(\cdot)$, 
Table \ref{tab:loaloacompare} a comparison of the point estimates of $p(\cdot)$, and Table \ref{tab:loaloacovparam} point
and interval estimates for 
the covariance parameters from each of the four methods. 
We find that \AGHQ{} is the closest to \MCMC{}, which illustrates the
accuracy of the \AGHQ{} results. 
The parameter $\rho$ is interpreted as the distance beyond which correlation between two observations
is practically negligible. Its estimate from \MCML{} is
far lower than for the Bayesian methods, which explains the different pattern in that map. Similarly, this 
estimate from \INLA{} is far higher than from \AGHQ{} and \MCMC{}, which explains the 
pattern in that map. The results from \AGHQ{} and \MCMC{} are similar to each other. It must be repeated that this only demonstrates the accuracy of \AGHQ{}, and does not suggest on its own that the others are inaccurate.

Table \ref{tab:loaloatiming} shows the runtimes and number of \MCMC{} iterations for the three non-\MCMC{} approaches.
Model fitting and spatial interpolation are separate steps, except for \INLA{} due to its use of the basis
function approximation. Here \AGHQ{} through \texttt{aghq::marginal\_laplace()} runs the slowest of the three non-sampling based methods, but still quite fast and, again, 
produces results much closer to
the \MCMC{} than the other two (with the above caveats).

It is not expected that \AGHQ{} should uniformly outperform 
these other well established approaches in this well studied example. Rather, the benefit of making inferences
with the \pkg{aghq} package is the flexibility to fit much more complicated models with little additional
code. This is demonstrated next, with an extension of this model to a setting where Bayesian
inferences had not previously been made.

\begin{table}[t]
\centering
\begin{tabular}{|l|rr|}
\hline
Method & $\text{mean}|p(\cdot) - p_{\MCMC{}}(\cdot)|$ & $\text{max}|p(\cdot) - p_{\MCMC{}}(\cdot)|$ \\
\hline
\AGHQ{} & 0.0139 & 0.101 \\
\INLA{} & 0.0304 & 0.225 \\
\MCML{} & 0.0463 & 0.218 \\
\hline
\end{tabular}
\caption{Average and maximum discrepancy with \MCMC{} in spatial interpolations for $p(\cdot)$ across the grid on which the maps
are drawn, for the three other appraches to inference for the \emph{Loa loa} example
of \S\ref{sec:loaloa}.}
\label{tab:loaloacompare}
\end{table}

\begin{table}[t]
\centering
\begin{tabular}{|l|rrr|rrr|}
\hline
Method & $2.5\%$ & $\hat{\sigma}$ & $97.5\%$ & $2.5\%$ & $\hat{\rho}$ & $97.5\%$ \\
\hline
\MCMC{} & 1.26 & 1.56 & 1.97 & 48.4 & 68.0 & 95.8 \\
\AGHQ{} & 1.20 & 1.48 & 1.81 & 45.4 & 63.9 & 87.5 \\
\INLA{} & 1.42 & 2.06 & 3.07 & 93.8 & 152. & 246. \\
\MCML{} & 1.23 & 1.40 & 1.60 & 12.3 & 15.7 & 19.9 \\
\hline
\end{tabular}
\caption{Point and interval estimares of covariance parameters $\sigma$ and $\rho$ (KM) for the four different appraches to inference for the \emph{Loa loa} example
of \S\ref{sec:loaloa}.}
\label{tab:loaloacovparam}
\end{table}

\begin{table}[t]
\centering
\begin{tabular}{|l|rr|rr|}
\hline
Method/Task & Fitting & Prediction & Total & Num. Iter \\
\hline
\MCMC{} & 465 & 2055 & 2520 & - \\
\AGHQ{} & 103 & 92.6 & 195 & 465 \\
\INLA{} & - & - & 42.7 & 102 \\
\MCML{} & 26.1 & 10.0 & 36.2 & 86.2 \\
\hline
\end{tabular}
\caption{Runtimes (seconds) and number of \MCMC{} iterations that could have been completed in the same time, for the four different appraches to inference for the \emph{Loa loa} example
of \S\ref{sec:loaloa}.}
\label{tab:loaloatiming}
\end{table}

\hypertarget{sec:loaloazip}{%
\subsection{Zero-inflated Geostatistical Binomial regression: extending
the Loa loa model}\label{sec:loaloazip}}

\citet{geostatlowresource} describe an extension of the geostatistical
binomial regression model which accounts for \emph{zero-inflation}: the notion that certain
spatial locations may be unsuitable for disease transmission and that this may lead
to more observed zero counts in the data than would be expected under the binomial model.
The likelihood is a discrete mixture of a binomial density and a point mass at zero, and 
both the transmission and mixing (\emph{suitibility}) probabilities depend on 
unknown spatial processes. The resulting model is not a latent Gaussian model and
is not compatible with \INLA{}. \citet{geostatlowresource} fit 
this challenging model to a different dataset using a frequentist maximum likelihood approach. 
However, this model is an \emph{extended} latent Gaussian model, and \citet{aghqus} fit it 
using approximate
Bayesian inference with \AGHQ{} using the \pkg{aghq} package, making use of its compatibility
with \pkg{TMB}. Here we demonstrate the details of the implementation of the zero inflated model using \pkg{aghq}, in the context of extending the non-zero inflated model of \S\ref{sec:loaloa}.

Let $\phi(\mb{s})$ denote the probability that location $\mb{s}\in\R^{2}$ is suitable
for disease transmission. The model
is: \begin{equation}\begin{aligned}
\mathbb{P}\left(Y_{i} = y_{i} | p(\mb{s}_{i}),\phi(\mb{s}_{i})\right) &= \left[1 - \phi(\mb{s}_{i})\right]\text{I}\left(y_{i} = 0\right) + \phi(\mb{s}_{i})\times\text{Binomial}[y_{i};N_{i},p(\mb{s}_{i})], \\
\log\left[\frac{p(\mb{s})}{1-p(\mb{s})}\right] &= \beta_{\texttt{risk}} + u(\boldsymbol{s}), \ \log\left[\frac{\phi(\mb{s})}{1-\phi(\mb{s})}\right] = \beta_{\texttt{zi}} + v(\boldsymbol{s}), \mb{s}\in\R^{2} \\
u(\cdot) | \boldsymbol{\theta} &\sim \mathcal{GP}(0,\text{C}_{\boldsymbol{\Xi},u}), \ v(\cdot) | \boldsymbol{\theta} \sim \mathcal{GP}(0,\text{C}_{\boldsymbol{\Xi},v}), \\
\end{aligned}\end{equation}
where $\bracevec{Y_{i}| p(\mb{s}_{i}),\phi(\mb{s}_{i}): i\in[n]}$ are independent as are $u(\cdot) | \boldsymbol{\theta}$ and $v(\cdot) | \boldsymbol{\theta}$. For simplicity the covariance parameters of the two spatial processes are constrained to be the same.

Making inferences from this model is very challenging due to the two competeing spatial processes, and not easily done using
any previously existing packages. To implement this model in \pkg{aghq}, the un-normalized
log-posterior $\log\pi(\mb{W},\mb{\theta},\mb{Y})$ is implemented in \pkg{TMB} with the 
Laplace approximation ``turned on'' for $\mb{W} = \left\{ u(\mb{s}_{i}), v(\mb{s}_{i}), i\in[n]; \beta_{\texttt{risk}}, \beta_{\texttt{zi}}\right\}$:
\begin{CodeChunk}
\begin{CodeInput}
R> ff <- TMB::MakeADFun(...,random = "W")
\end{CodeInput}
\end{CodeChunk}
The template \texttt{ff} has an element \texttt{ff\$fn()} which returns $-\log\widetilde{\pi}_{\LA}(\mb{\theta},\mb{Y})$,
including the necessary ``inner'' optimization of $\log\pi(\mb{W},\mb{\theta},\mb{Y})$ with respect to $\mb{W}$ (\S\ref{high-dimensions}). Further, 
an element \texttt{ff\$gr()} is provided which computes $-\partial_{\mb{\theta}}\log\widetilde{\pi}_{\LA}(\mb{\theta},\mb{Y})$
exactly and---critically---without re-optimizing over $\mb{W}$. This results in very efficient computations. This template
is then passed to \texttt{aghq::marginal\_laplace\_tmb()}:
\begin{CodeChunk}
\begin{CodeInput}
R> loaloazipquad <- marginal_laplace_tmb(
+     ff,3,startingvalue = c(paraminit$logkappa,paraminit$logtau)
+  )
\end{CodeInput}
\end{CodeChunk}
This returns an object identical to that returned by \texttt{aghq::marginal\_laplace()},
although usually runs significantly faster due to the efficiency of \pkg{TMB}. Note that only starting values
for $\mb{\theta}$ are provided to \pkg{aghq}, since the optimizations over $\mb{W}$ are handled entirely by \pkg{TMB}.
If different starting values for $\mb{W}$ are deemed necessary in a specific problem they can be passed to \texttt{TMB::MakeADFun()}
directly.

Inferences for this complicated model are then made in exactly the same manner as for the simpler model of \S\ref{sec:loaloa}.
Joint posterior samples are drawn from $\widetilde{\pi}(\mb{W}|\mb{Y})$:
\begin{CodeChunk}
\begin{CodeInput}
R> loazippostsamples <- sample_marginal(loaloazipquad,100)
\end{CodeInput}
\end{CodeChunk}
and then post-processed to compute estimates of any posterior functional of interest for $\mb{W}$. 

\begin{figure}[t]
\centering
\subfloat[$\phi(\cdot)$]{\includegraphics[width=.45\textwidth]{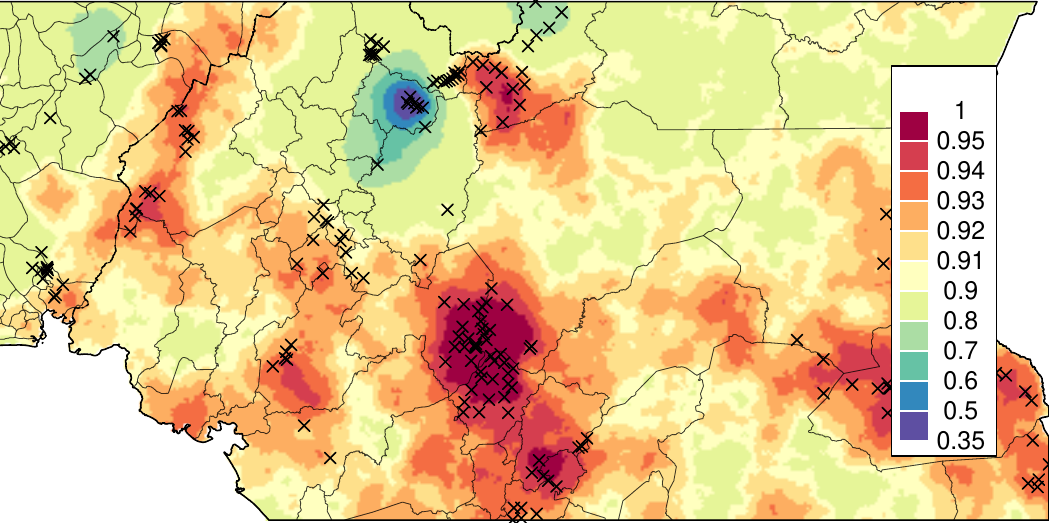}}
\subfloat[$\phi(\cdot)\times p(\cdot)$]{\includegraphics[width=.45\textwidth]{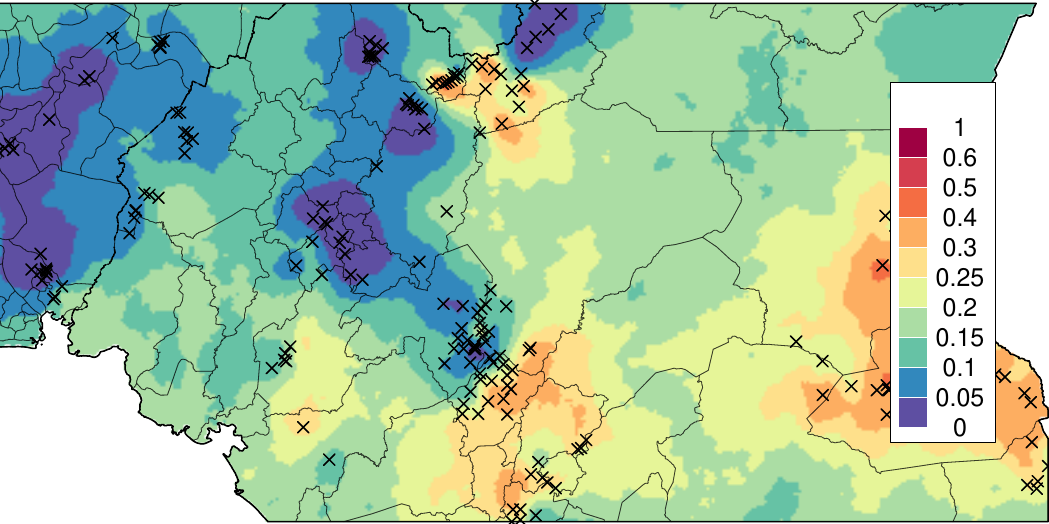}}
\caption{Spatial interpolations on a fine grid for posterior mean (a) suitability probabilities $\phi(\cdot)$ and 
(b) incidence probabilities $\phi(\cdot)\times p(\cdot)$ for the zero-inflated \emph{Loa loa} example
of \S\ref{sec:loaloazip}.}
\label{fig:loaloazipmaps}
\end{figure}

Figure \ref{fig:loaloazipmaps} shows the resulting maps of posterior mean suitability probabilities $\phi(\cdot)$ and
incidence probabilities $\phi(\cdot)\times p(\cdot)$, two complicated posterior functionals of $\mb{W}$.
The map of incidence is visually similar to that obtained with the non-zero inflated model of \S\ref{sec:loaloa}
(which effectively sets $\phi(\mb{s}) = 1$ for every $\mb{s}\in\R^{2}$). Novel to this example is the 
map of suitability in which it appears that a cluster of villages is identified which has low probability of being able
to transmit the disease.

\hypertarget{sec:glmmTMB}{%
\section{Leveraging Code from Existing Packages}\label{sec:glmmTMB}}

Many packages are available for fitting hierarchical models of the type (although
not always of the complexity) considered in \S\ref{sec:elgms}. Packages
which make approximate Bayesian inferences in these models are less common. The
\pkg{aghq} package, due to its flexibility, can be used as a wrapper for
making fast and accurate approximate Bayesian inferences in models which
other authors have written software to fit using other methods. For example,
\citet{disaggregation} make approximate Bayesian inferences in a very complicated
model for disease mapping using aggregated spatial point process data, and their
\pkg{disaggregation} package
creates a \pkg{TMB} template for the log-posterior and makes approximate 
Bayesian inferences using the corresponding Gaussian approximation provided by \pkg{TMB}. \citet{noeps}
make approximate Bayesian inferences for this model using the approximations
of \S\ref{high-dimensions}, yielding more accurate inferences (compared to \MCMC{}). Their
implementation simply involved taking the \pkg{TMB} function template constructed
using the \pkg{disaggregation} package to \texttt{aghq::marginal\_laplace\_tmb()},
and hence involved very little
additional code due to the flexibility of the \pkg{aghq} package.

The example in this section is from the \pkg{glmmTMB} package \citep{glmmTMB}, which 
provides a user-friendly interface for fitting a variety of complicated hierarchical models using Laplace-approximate maximum likelihood 
with the underlying computations done using \pkg{TMB}. The purpose of this example is
not necessarily to recommend that one method be used over the other, but rather to provide
further illustration of how approximate Bayesian inferences can be made in a straightforward
manner using \pkg{aghq} for users who have access to a \pkg{TMB} function template.

We consider their example of a zero-inflated, overdispersed Poisson
regression with random effects and where all of the mean, overdispersion, and
zero-inflation parameters depend on linear predictors. Counts \(Y_{i}, i\in[n]\) of salamanders in $n=23$
streams were measured 4 times per stream. Let $x_{\texttt{mined}}$ denote an indicator of 
whether a stream has been affected by mining, and $x_{\texttt{DOY}}$ denote the (centred and scaled)
day of the year. Further, let $d_{i} = 1$ if $\PP(Y_{i} = 0) = 1$ (a \emph{structural} zero) and $d_{i} = 0$ otherwise;
$p_{i} = \PP(d_{i} = 1)$; \(u_{i}\) a random intercept for site \(i\) with variance \(\sigma^{2}_{u}\); \(\lambda_{i} = \mathbb{E}(Y_{i}|u_{i},d_{i} = 0)\); and \(\sigma^{2}_{i} = \text{Var}(Y_{i}|\lambda_{i},\theta_{i},d_{i}=0)\). The model
is: \begin{equation}\begin{aligned}
Y_{i} | \lambda_{i},d_{i}=0,\sigma^{2}_{i} &\overset{ind}{\sim} \text{quasi-Poisson}(\lambda_{i},\sigma^{2}_{i}), \\
\log\lambda_{i} &= \beta_{0}^{\texttt{(mean)}} + \beta_{\texttt{mined}}\cdot x_{\texttt{mined},i} + u_{i}, \\
u_{i} &\overset{iid}{\sim}\text{Normal}(0,\sigma^{2}_{u}), \\
\sigma^{2}_{i} &= \lambda_{i}\left(1 + \frac{\lambda_{i}}{\theta_{i}}\right), \\
\log\theta_{i} &= \beta_{0}^{\texttt{(disp)}} + \beta_{\texttt{DOY}}\cdot x_{\texttt{DOY}}, \\
\log\left(\frac{p_{i}}{1-p_{i}}\right) &= \beta_{0}^{\texttt{(zi)}} + \beta_{\texttt{mined}}^{\texttt{(zi)}}\cdot x_{\texttt{mined},i}. \\
\end{aligned}\end{equation}
The model is fit using \pkg{glmmTMB} as follows:
\begin{CodeChunk}
\begin{CodeInput}
R> zipmod <- glmmTMB(count ~ mined + (1|site),zi = ~mined,
+    disp = ~DOY,data = Salamanders,
+    family = nbinom2,doFit = TRUE
+  ) 
\end{CodeInput}
\end{CodeChunk}
with a computation time of $3.14$ seconds.

To fit the model using \pkg{aghq}, obtain the data and parameter information by running 
the previous command but with \texttt{doFit=FALSE},
and then create the \pkg{TMB} template with the Laplace approximation turned on for $u$ and all the $\beta$
parameters:
\begin{CodeChunk}
\begin{CodeInput}
R> zipmodinfo <- glmmTMB(...,doFit=FALSE)
R> ff <- with(zipmodinfo,{
R>   TMB::MakeADFun(data = data.tmb,parameters = parameters,
+      random = names(parameters)[grep('theta',names(parameters),invert = TRUE)],
+      DLL = "glmmTMB",silent = TRUE
+    )
+  })
\end{CodeInput}
\end{CodeChunk}
The template \texttt{ff} implements $-\log\widetilde{\pi}_{\LA}(\psi,\mb{Y})$ where $\psi = \log\sigma_{u}$
(switching notation to be consistent with \citet{glmmTMB} who use $\theta$ for the overdispersion parameter).
The model is then fit using \pkg{aghq} in exactly the same manner as \S\ref{sec:loaloazip}:
\begin{CodeChunk}
\begin{CodeInput}
R> zipquad <- marginal_laplace_tmb(ff,3,0)
R> zipquadsamps <- sample_marginal(zipquad,1e03)
R> zipsigmapdf <- compute_pdf_and_cdf(
+    zipquad,list(totheta = log,fromtheta = exp),
+    finegrid = seq(-10,0.2,length.out = 1000)
+  )
\end{CodeInput}
\end{CodeChunk}
with a comparable computation time of $3.39$ seconds (for the above three tasks). A custom grid on which to approximate the density of $\sigma$ is provided, as the default
used by \pkg{aghq} was not wide enough to make the tail 
of the plot reach zero. This was done for aesthetic purposes only and does not change
any inferences for summary statistics.

\begin{figure}[t]
\centering
\makebox{
	\subfloat[$\widetilde{\pi}_{\LA}(\sigma_{u}|\mb{Y})$]{\includegraphics[width=.45\textwidth]{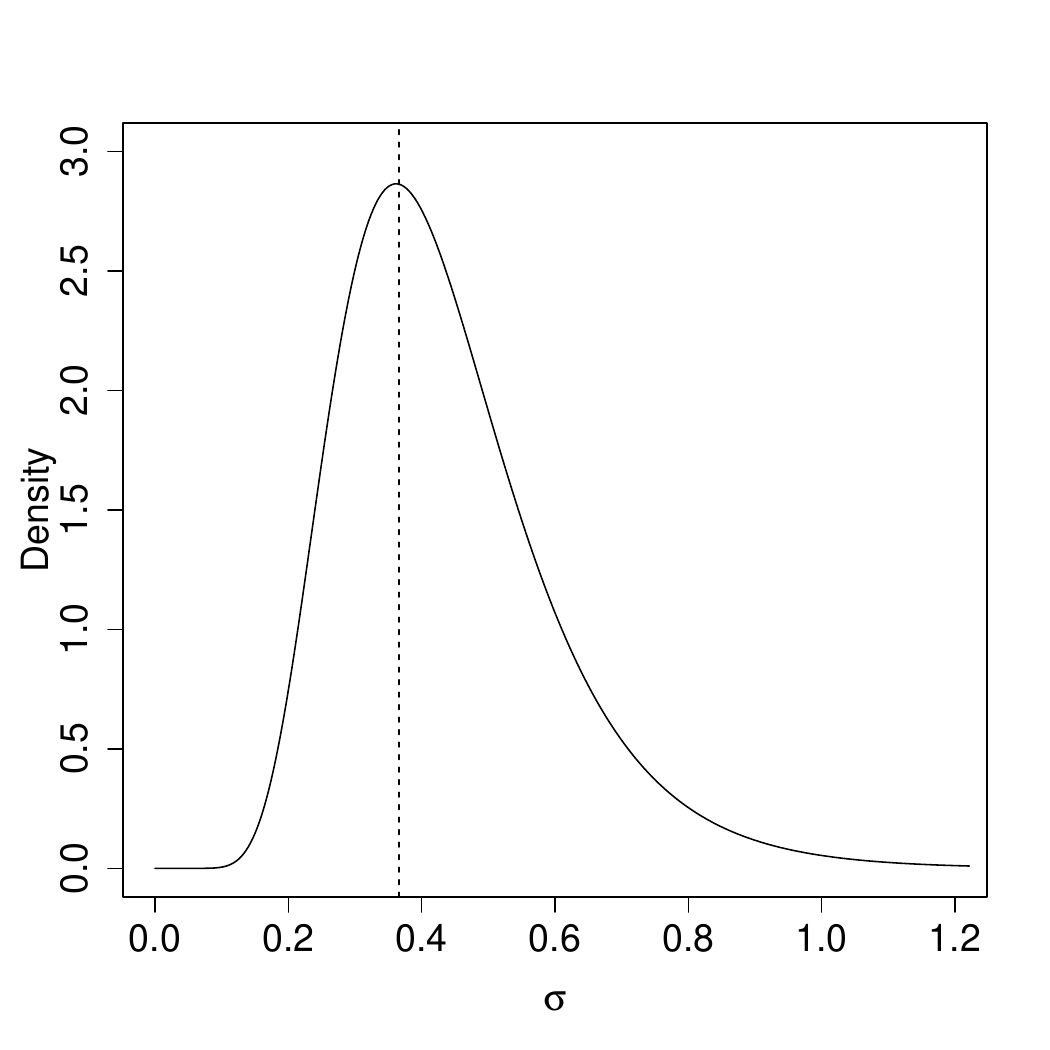}\label{fig:glmmtmbsigma}}
	\subfloat[$u_{i}$]{\includegraphics[width=.45\textwidth]{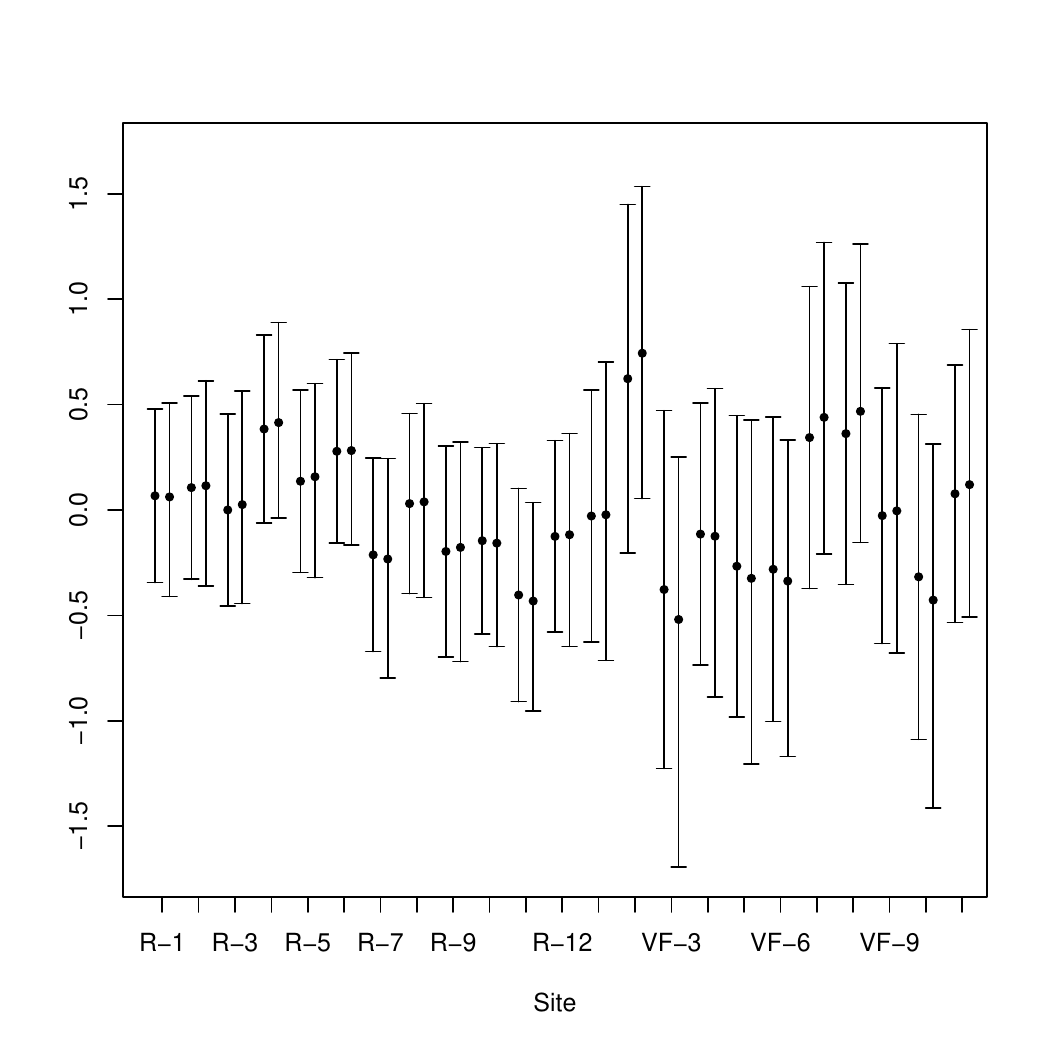}\label{fig:glmmtmbu}}
}
\caption{Comparison of inferences for random effects (a) standard deviation $\sigma_{u}$ and (b) estimates $u_{i}$ from \pkg{glmmTMB}
and \pkg{aghq} for the zero-inflated, overdispersed Poisson model of \S\ref{sec:glmmTMB}. The uncertainty in $\sigma_{u}$ is accounted for when making Bayesian inferences for $u_{i}$, resulting
in slightly wider credible intervals (right) than the corresponding confidence intervals from \pkg{glmmTMB} (left), although
the difference is not substantial in this example.
}
\label{fig:glmmtmbresults}
\end{figure}

Figure \ref{fig:glmmtmbsigma} shows a comparison of the approximate posterior for $\sigma_{u}$ from \pkg{aghq}
as well as from an \MCMC{} fit using \pkg{tmbstan}. The point estimate from \pkg{glmmTMB}
is included. We find that $\sigma_{u}$ is somewhat imprecisely estimated from these data and model, and hence it
is important to account for uncertainty in it when computing intervals for the random effects $u_{i}$.
Figure \ref{fig:glmmtmbu} shows the confidence and credible intervals for $u_{i},i\in[n]$ from \pkg{glmmTMB} and \pkg{aghq};
the latter are wider owing to the uncertainty induced by estimating $\sigma_{u}$, albeit only slightly in this example.

In principle, this approach could be used to make approximate Bayesian inferences in any model
fit using \pkg{glmmTMB}. A limitation is that a flat prior is 
used for $\psi$, and that there is no obvious way to change this without 
modifying the underlying \texttt{C++} template. However, doing so would defeat the purpose of using
the flexbile interface provided by \pkg{glmmTMB} to create the model. This strategy therefore
may not be appropriate in cases where the prior on $\psi$ is thought to be very important,
such as in the spatial models of \S\ref{sec:elgms}. Further, it was observed in this example
that the \MCMC{} results (not shown; code available) using \pkg{tmbstan} resulted in
somewhat different estimates of $\sigma$ than the two approaches which use the Laplace
approximation. It may be the case that neither \pkg{glmmTMB} nor \pkg{aghq} are producing
accurate results for this small-sample example, and this may be due to the Laplace approximation
being inaccurate. It would be of future interest to explore
alternative approximations to the Laplace approximation within \pkg{aghq}, 
such as those based on \AGHQ{} in moderate dimensions with sparse grids, as well as sequential reduction or importance sampling \citep{Ogden2013ASR}.

\hypertarget{sec:discussion}{%
\section{Discussion}\label{sec:discussion}}

The \pkg{aghq} package provides a flexible interface for making approximate Bayesian
inferences where the posterior is normalized using adaptive quadrature. Researchers who are implementing their own log-posteriors may use
\pkg{aghq} to make approximate Bayesian inferences in their models with very little
additional code. The \pkg{aghq} package is designed to work especially well with
models implemented in \pkg{TMB}, and can be used immediately to improve the accuracy of Bayesian inferences in any applications
where \pkg{TMB} is currently being used. In this manuscript, the \pkg{aghq} package has been demonstrated to provide fast
and quantifiably accurate approximate Bayesian inferences in several low-dimensional models. Further, the use of \pkg{aghq} has been demonstrated for fitting high-dimensional
latent Gaussian \citep{inla} and extended latent Gaussian \citep{lgmsplit,maxandsmooth,noeps} models for
making approximate Bayesian inferences in complicated non-Gaussian model-based geostatistical applications.

The \pkg{aghq} package is a ``mid-level'' package, aimed at specialist users who are implementing their
own log-posteriors. Its current impact is expected to be greater for
researchers who are developing new models than for researchers fitting existing models to
new data. Ongoing work involves using \pkg{aghq} as the basis for packages which implement 
approximate Bayesian inference for specific problems. Doing so involves creating a formula
interface to accept a model and data and create a \pkg{TMB} template (or otherwise), which is then
passed to \pkg{aghq} as described in \S\ref{sec:examples}, \S\ref{sec:elgms} and \S\ref{sec:glmmTMB}.
The \pkg{aghq} package automates all the tedious and routine---but highly nontrivial---computations
associated with making approximate Bayesian inferences, thereby allowing researchers to focus
more on the problem-specific challenges in each novel application.

\bibliography{biblio}

\begin{thebibliography}{35}
\newcommand{\enquote}[1]{``#1''}
\providecommand{\natexlab}[1]{#1}
\providecommand{\url}[1]{\texttt{#1}}
\providecommand{\urlprefix}{URL }
\expandafter\ifx\csname urlstyle\endcsname\relax
  \providecommand{\doi}[1]{doi:\discretionary{}{}{}#1}\else
  \providecommand{\doi}{doi:\discretionary{}{}{}\begingroup
  \urlstyle{rm}\Url}\fi
\providecommand{\eprint}[2][]{\url{#2}}

\bibitem[{Almutiry \emph{et~al.}(2020)Almutiry, V, and Deardon}]{epi}
Almutiry W, V VWK, Deardon R (2020).
\newblock \enquote{Continuous Time Individual-Level Models of Infectious
  Disease: EpiILMCT.}
\newblock \emph{arXiv:2006.00135v1}.

\bibitem[{Bates \emph{et~al.}(2015)Bates, M{\"a}chler, Bolker, and
  Walker}]{lme4}
Bates D, M{\"a}chler M, Bolker B, Walker S (2015).
\newblock \enquote{Fitting Linear Mixed-Effects Models Using {lme4}.}
\newblock \emph{Journal of Statistical Software}, \textbf{67}(1), 1--48.
\newblock \doi{10.18637/jss.v067.i01}.

\bibitem[{Bianconcini(2014)}]{aghqmle}
Bianconcini S (2014).
\newblock \enquote{Asymptotic properties of adaptive maximum likelihood
  estimators in latent variable models.}
\newblock \emph{Bernoulli}, \textbf{20}(3), 1507--1531.
\newblock \doi{10.3150/13-BEJ531}.

\bibitem[{Bilodeau \emph{et~al.}(2021)Bilodeau, Stringer, and Tang}]{aghqus}
Bilodeau B, Stringer A, Tang Y (2021).
\newblock \enquote{Stochastic Convergence Rates and Applications of Adaptive
  Quadrature in Bayesian Inference.}
\newblock \emph{arXiv:2102.06801 [stat.ME]}.

\bibitem[{Blocker(2018)}]{fastghquad}
Blocker AW (2018).
\newblock \emph{fastGHQuad: Fast 'Rcpp' Implementation of Gauss-Hermite
  Quadrature}.
\newblock R package version 1.0,
  \urlprefix\url{https://CRAN.R-project.org/package=fastGHQuad}.

\bibitem[{Brooks \emph{et~al.}(2017)Brooks, Kristensen, {van Benthem},
  Magnusson, Berg, Nielsen, Skaug, Maechler, and Bolker}]{glmmTMB}
Brooks ME, Kristensen K, {van Benthem} KJ, Magnusson A, Berg CW, Nielsen A,
  Skaug HJ, Maechler M, Bolker BM (2017).
\newblock \enquote{{glmmTMB} Balances Speed and Flexibility Among Packages for
  Zero-inflated Generalized Linear Mixed Modeling.}
\newblock \emph{The R Journal}, \textbf{9}(2), 378--400.
\newblock
  \urlprefix\url{https://journal.r-project.org/archive/2017/RJ-2017-066/index.html}.

\bibitem[{Brown(2011)}]{geostatsp}
Brown P (2011).
\newblock \enquote{Model-based geostatistics the easy way.}
\newblock \emph{Journal of Statistical Software}, \textbf{73}(4), 423--498.

\bibitem[{Davis and Rabinowitz(1975)}]{numint}
Davis PJ, Rabinowitz P (1975).
\newblock \emph{Methods of Numerical Integration}.
\newblock Academic Press.

\bibitem[{Diggle and Ribeiro(2006)}]{loaloa}
Diggle P, Ribeiro P (2006).
\newblock \emph{Model-Based Geostatistics}.
\newblock Springer-Verlag.

\bibitem[{Diggle and Giorgi(2016)}]{geostatlowresource}
Diggle PJ, Giorgi E (2016).
\newblock \enquote{Model-Based Geostatistics for Prevalence Mapping in
  Low-Resource Settings.}
\newblock \emph{Journal of the American Statistical Association},
  \textbf{111}(515), 1096--1120.

\bibitem[{Eadie and Harris(2016)}]{gwen2}
Eadie GM, Harris WE (2016).
\newblock \enquote{Bayesian mass estimates of the Milky Way: the dark and light
  sides of parameter assumptions.}
\newblock \emph{The Astrophysical Journal}, \textbf{829}(108).

\bibitem[{Evans and Swartz(2000)}]{evansintegrations}
Evans M, Swartz T (2000).
\newblock \emph{Approximating Integrals via Monte Carlo and Deterministic
  Methods}.
\newblock Oxford University Press.

\bibitem[{Fuglstad \emph{et~al.}(2019)Fuglstad, Simpson, Lindgren, and
  Rue}]{pcpriormatern}
Fuglstad GA, Simpson D, Lindgren F, Rue H (2019).
\newblock \enquote{Constructing priors that penalize the complexit of Gaussian
  random fields.}
\newblock \emph{Journal of the American Statistical Association},
  \textbf{114}(525), 445--452.

\bibitem[{Gerstner and Griebel(1998)}]{sparsegrids}
Gerstner T, Griebel M (1998).
\newblock \enquote{Numerical Integration Using Sparse Grids.}
\newblock \emph{Numerical Algorithms}, \textbf{18}(32), 209--.

\bibitem[{Giorgi and Diggle(2017)}]{prevmap}
Giorgi E, Diggle P (2017).
\newblock \enquote{{PrevMap}: An R Package {for} Prevalence Mapping.}
\newblock \emph{Journal of Statistical Software}, \textbf{78}.
\newblock \doi{10.18637/jss.v078.i08}.

\bibitem[{Hrafnkelsson \emph{et~al.}(2021)Hrafnkelsson, Siegert, Huser, Bakka,
  and Árni V.~Jóhannesson}]{maxandsmooth}
Hrafnkelsson B, Siegert S, Huser R, Bakka H, Árni V~Jóhannesson (2021).
\newblock \enquote{{Max-and-Smooth: A Two-Step Approach for Approximate
  Bayesian Inference in Latent Gaussian Models}.}
\newblock \emph{Bayesian Analysis}, \textbf{16}(2), 611 -- 638.
\newblock \doi{10.1214/20-BA1219}.
\newblock \urlprefix\url{https://doi.org/10.1214/20-BA1219}.

\bibitem[{Jin and Andersson(2020)}]{adaptive_GH_2020}
Jin S, Andersson B (2020).
\newblock \enquote{{A note on the accuracy of adaptive Gauss–Hermite
  quadrature}.}
\newblock \emph{Biometrika}.

\bibitem[{Keshavarzzadeh \emph{et~al.}(2018)Keshavarzzadeh, Kirby, and
  Narayan}]{designedquadrature}
Keshavarzzadeh V, Kirby RM, Narayan A (2018).
\newblock \enquote{Numerical Integration in Multiple Dimensions with Designed
  Quadrature.}
\newblock \emph{SIAM Journal of Scientific Computing}, \textbf{40}(4),
  A2033--A2061.

\bibitem[{Kristensen \emph{et~al.}(2016)Kristensen, Nielson, Berg, Skaug, and
  Bell}]{tmb}
Kristensen K, Nielson A, Berg CW, Skaug H, Bell BM (2016).
\newblock \enquote{TMB: automatic differentiation and Laplace approximation.}
\newblock \emph{Journal of statistical software}, \textbf{70}(5).

\bibitem[{Lindgren and Rue(2015)}]{inlasoftware}
Lindgren F, Rue H (2015).
\newblock \enquote{{Bayesian spatial modelling with {R}-{INLA}}.}
\newblock \emph{Journal of Statistical Software}, \textbf{63}(19), 1--25.
\newblock \urlprefix\url{http://www.jstatsoft.org/v63/i19/}.

\bibitem[{Lindgren \emph{et~al.}(2011)Lindgren, Rue, and Lindstr{\"o}m}]{spde}
Lindgren F, Rue H, Lindstr{\"o}m J (2011).
\newblock \enquote{An explicit link between Gaussian fields and Gaussian Markov
  random fields: the stochastic partial differential equation approach.}
\newblock \emph{Journal of the Royal Statistical Society, Series B (Statistical
  Methodology)}, \textbf{73}(4), 423--498.

\bibitem[{Liu and Pierce(1994)}]{adaptive_GH_1994}
Liu Q, Pierce DA (1994).
\newblock \enquote{{A note on Gauss-Hermite quadrature}.}
\newblock \emph{Biometrika}, \textbf{81}(3), 624--629.

\bibitem[{Monnahan and Kristensen(2018)}]{tmbstan}
Monnahan CC, Kristensen K (2018).
\newblock \enquote{No-U-turn sampling for fast Bayesian inference in ADMB and
  TMB: Introducing the adnuts and tmbstan R packages.}
\newblock \emph{PLOS One}, \textbf{13}(5).
\newblock \doi{https://doi.org/10.1371/journal.pone.0197954}.

\bibitem[{Nandi \emph{et~al.}(2020)Nandi, Lucas, Arambepola, Gething, and
  Weiss}]{disaggregation}
Nandi AK, Lucas TCD, Arambepola R, Gething P, Weiss D (2020).
\newblock \enquote{{disaggregation: an R package for Bayesian spatial
  disaggregation modelling}.}
\newblock \emph{arxiv}.
\newblock \urlprefix\url{https://arxiv.org/abs/2001.04847}.

\bibitem[{Naylor and Smith(1982)}]{nayloradaptive}
Naylor J, Smith AFM (1982).
\newblock \enquote{Applications of a Method for the Efficient Computation of
  Posterior Distributions.}
\newblock \emph{Journal of the Royal Statistical Society, Series C (Applied
  Statistics)}, \textbf{31}(3), 214--225.

\bibitem[{Ogden(2013)}]{Ogden2013ASR}
Ogden H (2013).
\newblock \enquote{A sequential reduction method for inference in generalized
  linear mixed models.}
\newblock \emph{arXiv: Computation}.

\bibitem[{Rizopoulos(2020)}]{glmmadaptive}
Rizopoulos D (2020).
\newblock \emph{GLMMadaptive: Generalized Linear Mixed Models using Adaptive
  Gaussian Quadrature}.
\newblock R package version 0.7-15,
  \urlprefix\url{https://CRAN.R-project.org/package=GLMMadaptive}.

\bibitem[{Rue \emph{et~al.}(2009)Rue, Martino, and Chopin}]{inla}
Rue H, Martino S, Chopin N (2009).
\newblock \enquote{{Approximate Bayesian inference for latent Gaussian models
  by using integrated nested Laplace approximations}.}
\newblock \emph{Journal of the Royal Statistical Society. Series B (Statistical
  Methodology)}, \textbf{71}(2), 319--392.

\bibitem[{Schlather \emph{et~al.}(2015)Schlather, Malinowski, Menck, Oesting,
  and Strokorb}]{randomfields}
Schlather M, Malinowski A, Menck PJ, Oesting M, Strokorb K (2015).
\newblock \enquote{Analysis, simulation and prediction of multivariate random
  fields with package RandomFields.}
\newblock \emph{Journal of Statistical Software}, \textbf{63}(8).

\bibitem[{Stringer \emph{et~al.}(2020)Stringer, Brown, and
  Stafford}]{casecrossover}
Stringer A, Brown P, Stafford J (2020).
\newblock \enquote{Approximate {B}ayesian Inference for Case Crossover Models.}
\newblock \emph{Biometrics (To appear)}.

\bibitem[{Stringer \emph{et~al.}(2021)Stringer, Brown, and Stafford}]{noeps}
Stringer A, Brown P, Stafford J (2021).
\newblock \enquote{{Fast, Scalable Approximations to Posterior Distributions in
  Extended Latent Gaussian Models}.}
\newblock \emph{arXiv:2103.07425 [stat.ME]}.

\bibitem[{Tierney and Kadane(1986)}]{laplace}
Tierney L, Kadane JB (1986).
\newblock \enquote{{Accurate approximations for posterior moments and marginal
  densities}.}
\newblock \emph{Journal of the American Statistical Association},
  \textbf{81}(393), 82--86.

\bibitem[{Villandré \emph{et~al.}(2020)Villandré, Plante, Duchesne, and
  Brown}]{inlamra}
Villandré L, Plante JF, Duchesne T, Brown P (2020).
\newblock \enquote{{INLA-MRA: a Bayesian method for large spatiotemporal
  datasets}.}
\newblock \emph{arXiv:2004.10101 [stat.CO]}.

\bibitem[{Wachter and Biegler(2006)}]{ipopt}
Wachter A, Biegler LT (2006).
\newblock \enquote{On the Implementation of a Primal-Dual Interior Point Filter
  Line Search Algorithm for Large-Scale Nonlinear Programming.}
\newblock \emph{Mathematical Programming}, \textbf{106}(1), 25--57.

\bibitem[{Óli Páll~Geirsson \emph{et~al.}(2020)Óli Páll~Geirsson,
  Hrafnkelsson, Simpson, and Sigurdarson}]{lgmsplit}
Óli Páll~Geirsson, Hrafnkelsson B, Simpson D, Sigurdarson H (2020).
\newblock \enquote{{LGM Split Sampler: An Efficient MCMC Sampling Scheme for
  Latent Gaussian Models}.}
\newblock \emph{Statistical Science}, \textbf{35}(2), 218 -- 233.
\newblock \doi{10.1214/19-STS727}.
\newblock \urlprefix\url{https://doi.org/10.1214/19-STS727}.

\end{thebibliography}

\end{document}